\documentclass[iop,twocolumn]{emulateapj}

\usepackage[breaklinks,colorlinks=true,linkcolor=blue,anchorcolor=blue,citecolor=blue,urlcolor=blue,draft=false]{hyperref}
\usepackage{color}
\usepackage{aas_macros}
\usepackage{amssymb,amsmath}
\usepackage{etoolbox}
\usepackage{natbib}

\newcommand{\simgt}{\lower.5ex\hbox{$\; \buildrel > \over \sim \;$}}
\newcommand{\simlt}{\lower.5ex\hbox{$\; \buildrel < \over \sim \;$}}

\def\btheta{\mbox{\boldmath $\theta$}}

\def\munit{\mbox{$10^{15}\,M_\sun$}}

\def\dlsds{\mbox{$D_\mathrm{ls}/D_\mathrm{s}$}}
\def\RC{\mbox{$R_\mathrm{C}$}}

    \makeatletter
    
    \patchcmd{\NAT@citex}
      {\@citea\NAT@hyper@{%
         \NAT@nmfmt{\NAT@nm}%
         \hyper@natlinkbreak{\NAT@aysep\NAT@spacechar}{\@citeb\@extra@b@citeb}%
         \NAT@date}}
      {\@citea\NAT@nmfmt{\NAT@nm}%
       \NAT@aysep\NAT@spacechar\NAT@hyper@{\NAT@date}}{}{}
    
    \patchcmd{\NAT@citex}
      {\@citea\NAT@hyper@{%
         \NAT@nmfmt{\NAT@nm}%
         \hyper@natlinkbreak{\NAT@spacechar\NAT@@open\if*#1*\else#1\NAT@spacechar\fi}%
           {\@citeb\@extra@b@citeb}%
         \NAT@date}}
      {\@citea\NAT@nmfmt{\NAT@nm}%
       \NAT@spacechar\NAT@@open\if*#1*\else#1\NAT@spacechar\fi\NAT@hyper@{\NAT@date}}
      {}{}
    
    \makeatother

\def\equationautorefname~#1\null{Equation~(#1)\null}

\begin{document}
\hypersetup{pdfauthor={Elinor Medezinski}, pdftitle={FF: Subaru WL
    Analysis of the Merging Galaxy Cluster A2744}}
\title{Frontier Fields: Subaru Weak-Lensing Analysis of the Merging Galaxy Cluster A2744}

\author{Elinor Medezinski\altaffilmark{1,2,3}}      %
\author{Keiichi Umetsu\altaffilmark{4}}  
\author{Nobuhiro Okabe\altaffilmark{5,6,7}} 
\author{Mario Nonino\altaffilmark{8}}           
\author{Sandor Molnar\altaffilmark{9,10}}           
\author{Richard Massey\altaffilmark{11,12}}           
\author{Renato Dupke\altaffilmark{13,14,15}}           
\author{Julian Merten\altaffilmark{16}}           

\email{elinorm@astro.princeton.edu}
\altaffiltext{*}{Based in part on data collected at the Subaru Telescope,
  which is operated by the National Astronomical Society of Japan.}
\altaffiltext{1}{Department of Astrophysical Sciences, 4 Ivy Lane, Princeton, NJ 08544, USA} 
\altaffiltext{2}{Center for Astrophysics and Planetary Science, Racah Institute of Physics, The Hebrew University, Jerusalem 91904, Israel} 
\altaffiltext{3}{Department of Physics and Astronomy, The Johns Hopkins
University, 3400 North Charles Street, Baltimore, MD 21218, USA} 

\altaffiltext{4}{Institute of Astronomy and Astrophysics, Academia
Sinica, P.~O. Box 23-141, Taipei 10617, Taiwan.}  
\altaffiltext{5}{Department of Physical Science, Hiroshima University, 1-3-1 Kagamiyama, Higashi-Hiroshima, Hiroshima 739-8526, Japan}
\altaffiltext{6}{Hiroshima Astrophysical Science Center, Hiroshima University, Higashi-Hiroshima, Kagamiyama 1-3-1, 739-8526, Japan}
\altaffiltext{7}{Kavli Institute for the Physics and Mathematics of the Universe (WPI), Todai Institutes for Advanced Study,University of Tokyo, 5-1-5 Kashiwanoha, Kashiwa, Chiba 277-8583, Japan}
\altaffiltext{8}{INAF/Osservatorio Astronomico di Trieste, via
G.B. Tiepolo 11, 34143 Trieste, Italy}
\altaffiltext{9}{WinSet, Falls Church, 6107E Arlington Blvd., 22044 VA}
\altaffiltext{10}{Department of Physics, National Taiwan University, Taipei 10617, Taiwan}
\altaffiltext{11}{Institute for Computational Cosmology, Durham University, South Road, Durham DH1 3LE, UK}
\altaffiltext{12}{Centre for Extragalactic Astronomy, Durham University, South Road, Durham DH1 3LE, UK}
\altaffiltext{13}{Observat\'orio Nacional, Rua Gal. J\'ose Cristino, 20921-400 Rio de Janeiro, Brazil}
\altaffiltext{14}{University of Michigan, Ann Arbor MI 48109, USA}
\altaffiltext{15}{Eureka Scientific Inc., 2452 Delmer St. Suite 100, Oakland, CA
94602, USA}
\altaffiltext{16}{Department of Physics, University of Oxford, Keble Road, Oxford OX1 3RH, UK}

\begin{abstract}
We present a weak-lensing analysis of the merging {\em Frontier Fields} (FF) cluster Abell~2744 using new Subaru/Suprime-Cam imaging. The wide-field lensing mass distribution  reveals this cluster is comprised of four distinct substructures.  Simultaneously modeling the two-dimensional reduced shear field using a combination of a Navarro--Frenk--White (NFW) model for the main core and truncated NFW models for the subhalos, we determine their masses and locations. The total mass of the system is constrained as $M_\mathrm{200c} = (2.06\pm0.42)\times\munit$.  The most massive clump is the southern component with $M_\mathrm{200c} = (7.7\pm3.4)\times10^{14}\,M_\odot$, followed by the western substructure  ($M_\mathrm{200c} = (4.5\pm2.0)\times10^{14}\,M_\odot$) and two smaller substructures to the northeast ($M_\mathrm{200c} = (2.8\pm1.6)\times10^{14}\,M_\odot$) and northwest ($M_\mathrm{200c} = (1.9\pm1.2)\times10^{14}\,M_\odot$). The presence of the four substructures supports the picture of multiple mergers. Using a composite of hydrodynamical binary simulations we explain this complicated system without the need for a ``slingshot'' effect to produce the northwest X-ray interloper, as previously proposed. The locations of the substructures appear to be offset from both the gas ($87^{+34}_{-28}$ arcsec, 90\% CL) and  the galaxies ($72^{+34}_{-53}$ arcsec, 90\% CL) in the case of the northwestern and western subhalos. To confirm or refute these findings, high resolution space-based observations extending beyond the current FF limited coverage to the west and northwestern area are essential.
\end{abstract}
 

\section{Introduction} 
\label{sec:intro}

Clusters of galaxies depict the most recent bound phase of the hierarchical structure formation, evolving and relaxing by accreting matter along large scale filaments. In some extreme cases, clusters are caught in the act of violent head-on collisions between groups or clusters of similar masses.  A key tool in understanding the physics governing such events is the comparison between the distributions of the three main components involved -- the dark matter (DM), the hot intracluster gas and the stars in galaxies.  

Although it comprises 85\% of all the matter in the Universe \citep{2013ApJS..208...19H}, the nature of DM is still largely unknown. 
It is thought to be cold and collisionless like the galaxies. 
The intracluster gas, on the other hand, is highly dissipative and can, in extreme cases, be separated from the DM and galaxies by ram-pressure stripping. The best example to date of such stripping is seen in the ``Bullet'' cluster \citep{Markevitch02}, a high-velocity merger viewed in the plane of the sky. The gas of the bullet, as seen by X-ray, is lagging behind the DM component, deduced from weak gravitational lensing (WL) \citep{Clowe04}.
Measuring the displacement of the X-ray gas peaks from the gravitational mass peaks in this cluster provided the first upper limit on the cross-section for collision of DM, $<1\, \mathrm{cm}^2\, \mathrm{g}^{-1}$ \citep{Markevitch2004}.

Due to their enormous gravitation potential, clusters of galaxies also
act as powerful cosmic lenses, enhancing light coming from galaxies that
formed in the early universe. Recent discoveries were made of galaxies
out to redshifts as high as $z\sim11$ \citep{Zheng2012,Coe:2013qy},
being magnified by massive clusters observed as part of CLASH, a
multi-cycle treasury program with the {\em Hubble Space Telescope
(HST)}\citep{Postman+2012_CLASH}. This inspired the ongoing Frontier
Fields\footnote{\url{http://www.stsci.edu/hst/campaigns/frontier-fields/}} 
(FF) initiative \citep{Lotz2014}, an ambitious {\em HST}  program to
deeply observe six strong-lens (SL) clusters and detect high-redshift
galaxies. The legacy FF program will fortuitously provide deep,
high-resolution imaging of the cluster cores. Accurate maps of the mass distribution in these clusters are needed in order to quantify the  magnification of the distant galaxies and measure the star formation history in the early universe.

Most of the FF clusters are known to be non-relaxed systems. One of these clusters,  Abell~2744 (hereafter A2744; $z=0.308$) \citep{Lotz2013}, shows one of the most complicated merger phenomena to have been detected. It was first recognized as a merger by the presence of a luminous radio halo and a radio relic \citep{Giovannini1999,Govoni2001,Govoni2001a}. {\it Chandra} X-ray studies  revealed intricate substructure \citep{Kempner2004,Owers2011}, including a prominent hot gas cloud situated between the two main galaxy partitions,  cold and dense remnant cores to the south and north, tidal debris between them, and an interloping cloud of gas to the north-west, with no associated galaxies. These all pointed to a north-south merger with roughly equal mass, and a further dynamical study indicated there is a large line-of-sight (LOS) component to the merger axis \citep{Boschin2006}. However, it did not give a clear picture where lies the massive cluster core, to the north \citep{Owers2011} or to the south \citep{Kempner2004}, and whether the northwest X-ray interloper was falling in or had already passed from the southeast.
A recent lensing analysis by \citep[][hereafter M11]{Merten11}, mainly using {\it HST} and VLT imaging data, showed that the main cluster potential is situated near the southern part, but confounded the picture even more by finding 
a ``dark core'' of DM-only material between the two gas clouds, whereas the X-ray interloper seemed to be empty of DM as well as galaxies. M11 suggested a complicated ``slingshot'' scenario where the gas was thrown past the galaxies and DM, to try and explain this unclear picture.

In this paper, we perform an improved wide-field WL analysis of the cluster A2744, using new deep Subaru/Suprime-Cam observations that cover the full extent ($\lesssim5$\,Mpc) of this intermediate redshift cluster. Our multi-band imaging allows us to obtain a reconstructed cluster total mass distribution free of dilution by foreground structures, a main source of systematics in WL studies. We aim to detect and accurately constrain the physical properties of the different substructures of this merging cluster, so that a clear picture of the merger scenario can be illustrated by custom-made hydrodynamical simulations. 

This paper is organized as follows.   In \autoref{sec:theory} we briefly summarize the basic theory of WL. In \autoref{sec:data} we present the observation, their reduction and processing. In \autoref{sec:WL} we present the WL analysis, including the mass reconstruction, substructure detection and multi-halo modeling of their masses.  In \autoref{sec:discussion} we discuss our results, compare with earlier studies and  revisit the interpretation of the merger scenario in light of our findings. We finally summarize in \autoref{sec:summary}.
Throughout this paper, we use the AB magnitude system,
and adopt a concordance $\Lambda$CDM cosmology with $\Omega_m =0.3$,
$\Omega_\Lambda =0.7$, and $H_0 =100\,h$\,km\,s$^{-1}$\,Mpc$^{-1}$ with $h\equiv0.7h_{70}=0.7$. In
this cosmology, $1\arcmin$ corresponds to 272\,kpc at
the cluster redshift, $z = 0.308$. 
The cluster center used is
R.A.$=$00:14:18.9, Decl.$=-$30:23:22 (J2000.0) \citep{Ebeling10}, which closely corresponds to the cluster X-ray brightness peak.

\section{WL Methodology}
\label{sec:theory}

In this work, we infer from lensing the surface mass density, a.k.a. convergence, $\kappa(\btheta) =
\Sigma(\btheta)/\Sigma_{\mathrm crit}$, which is expressed in units
of the critical surface-mass density for lensing,  $\Sigma_{\mathrm crit}  =
(c^2 D_\mathrm{s})/(4\pi G D_\mathrm{l}D_\mathrm{ls})
\equiv c^2/(4\pi G D_\mathrm{l}\beta)$,
where $\beta(z_\mathrm{s},z_\mathrm{l}) \equiv \dlsds$ is the lensing
``depth'' for a source at redshift $z_\mathrm{s}$, 
and $D_\mathrm{l},\,D_\mathrm{s},$ and $D_\mathrm{ls}$ are the lens, source, and lens--source 
angular diameter distances, respectively. 

The shear field induced by gravitational lensing,
$\gamma=\gamma_1+i\gamma_2$, is non-locally related to the convergence
through 
\begin{equation}
 \gamma(\btheta) =  \frac{1}{\pi}\int d^2\theta' D(\btheta-\btheta') \kappa(\btheta')
\end{equation}
with the kernel being $D(\btheta) = (\theta_2^2-\theta_1^2-2i\theta_1\theta_2)/|\btheta|^4$ \citep{KaiserSquires93}. The convolution theorem yields in turn the convergence,
\begin{equation}
\label{eq:KS93}
 \kappa(\btheta) = \frac{1}{\pi} \int d^2\theta' D^*(\btheta-\btheta') \gamma(\btheta')
\end{equation}

In the weak regime, the reduced shear \citep[see, e.g.,][]{Bartelmann01},
\begin{equation}
\label{eq:g}
g(\btheta) \equiv \gamma(\btheta)/(1-\kappa(\btheta)),
\end{equation}
is the actual observable, derived from the observed ellipticities of background galaxies \citep{KSB}.
We calculate the weighted average of reduced shear 
on a pixelized Cartesian grid
($m=1,2,...,N_\mathrm{pix}$) as
\begin{equation}
\label{eq:ggrid}
 \langle g(\btheta_m)\rangle = \frac{\sum_{i} S(\theta_i,\theta_m) w_i g_i}{\sum_{i} S(\theta_i,\theta_m) w_i}
\end{equation}
where $S(\theta_i,\theta_m)$ is a spatial window function, $g_i$ is the estimate for the reduced shear of the $i$th object at $\btheta_i$, and  $w_i$ is the statistical  weight for the $i$th object, given by
\begin{equation}
\label{eq:w}
 w_i=1/(\sigma_{g,i}^2+\alpha^2),
\end{equation}
where $\sigma_{g,i}^2$ is the error variance of $g_i$, and $\alpha^2$ is a softening constant variance. We set $\alpha=0.4$, which is a typical value of the mean RMS $\bar{\sigma}_g$ found in Subaru
observations \citep[e.g.,][]{UB2008,Okabe2010}.
The variance on the pixelized shear map is then given by \citet{Umetsu+2009,Umetsu2015}
\begin{equation}
\label{eq:siggrid}
 \sigma_{g}^2(\btheta_m) = \frac{\sum_{i} S^2(\theta_i,\theta_m) w^2_i \sigma_{g,i}^2}{(\sum_{i} S(\theta_i,\theta_m) w_i)^2}.
\end{equation}

\section{Subaru Observations}
\label{sec:data}

In this section we present the
data reduction and analysis of A2744 based on deep 
multi-color imaging (\autoref{subsec:data}).  
We briefly describe our WL shape measurement procedure in \autoref{subsec:shape}.
The galaxy sample selection is described in \autoref{subsec:samples} and the samples depth estimation is given in \autoref{subsec:depth}.
\subsection{Data reduction and Photometry}
\label{subsec:data}


\begin{deluxetable}{ccccc}
\tablecolumns{5}
\tablecaption{
 \label{tab:data}
Optical Imaging Data
} 
\tablewidth{0pt} 
\tablehead{ 
 \multicolumn{1}{c}{Instrument} &
 \multicolumn{1}{c}{Filter} &
 \multicolumn{1}{c}{Exposure time} &
 \multicolumn{1}{c}{Seeing} &
 \multicolumn{1}{c}{$m_{\mathrm lim}$\tablenotemark{a}}  
\\
 \colhead{} &
 \colhead{} &
 \multicolumn{1}{c}{(ks)} &
 \multicolumn{1}{c}{(arcsec)} &
 \multicolumn{1}{c}{(AB mag)} 
} 
\startdata  
Subaru/S-cam & $B$  & 2.1 & 1.08 & 27.58\\
ESO/WFI  & $V$  & 2.7 & 0.9 & 25.59\\
Subaru/S-cam & $\RC$\tablenotemark{b}  & 3.12 & 1.16 & 26.83 \\ 
Subaru/S-cam & $i'$  & 1.68 & 1.42 & 26.31 \\
Subaru/S-cam & $z'$  & 3.6 & 1.07 & 26.03 \\
VLT/HAWK-I\tablenotemark{c}& $K_{\mathrm S}$  & 47.52 & 0.4 & 25.25 
\enddata
\tablenotetext{a}{Limiting magnitude for a $3\sigma$ detection within a
 $2\arcsec$ aperture.}
\tablenotetext{b}{Band used for WL shape measurements.}
\tablenotetext{c}{Instrument only covers central $4\arcmin$.}
\end{deluxetable}

\begin{figure*}[!htb]
 \begin{center}
\includegraphics[width=0.48\textwidth,clip]{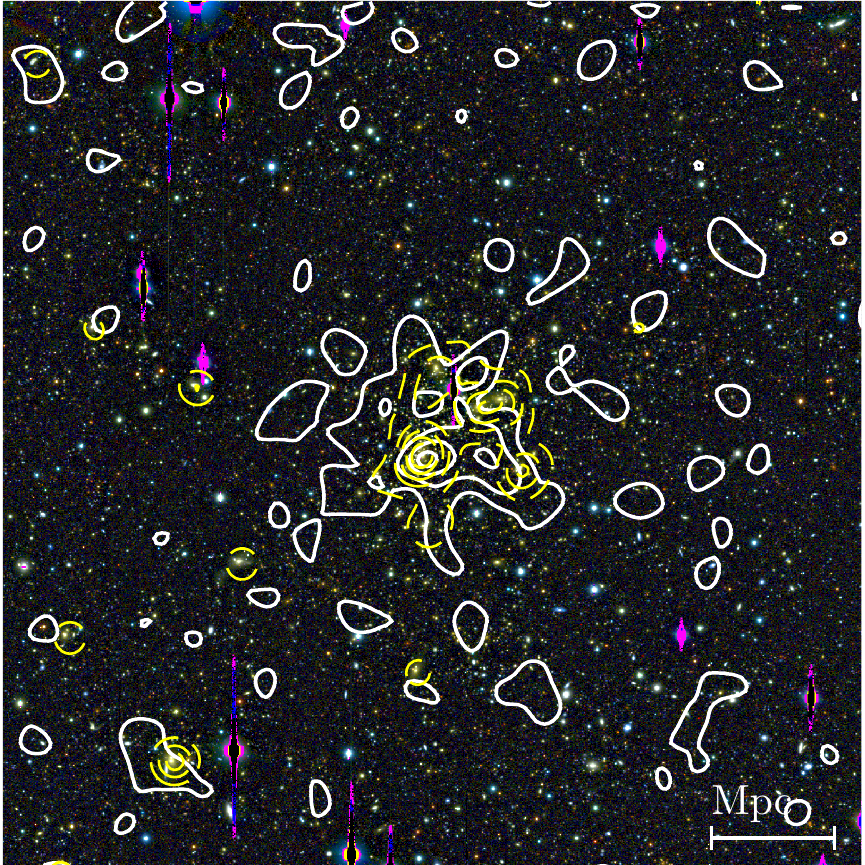}
\includegraphics[width=0.48\textwidth,clip]{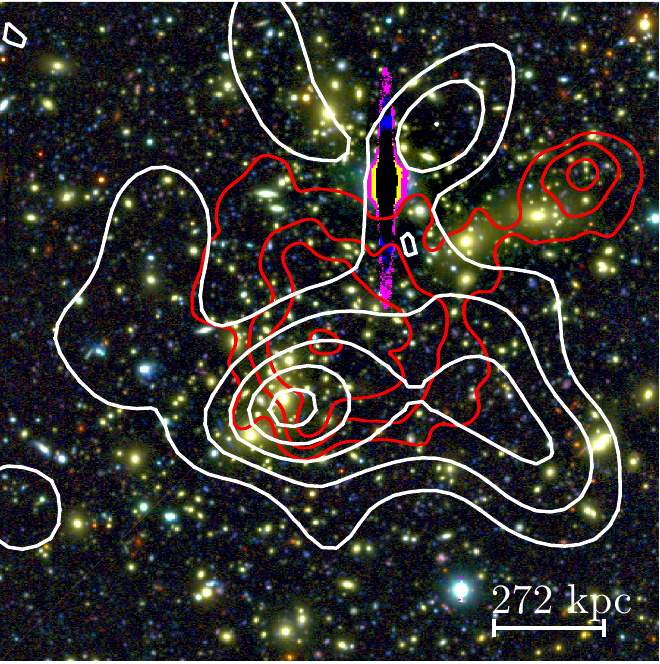}
\caption{ {\it Left:}
$26\arcmin\times26\arcmin$ Subaru $BR_{\mathrm C}z'$ false color
image of A2744 ($z=0.308$). We overlay  the surface
mass density map reconstructed from the Subaru WL analysis (white, starting at $1.5\sigma_\kappa$, in $3\sigma_\kappa$ increments), and
the luminosity density map  (yellow dashed).
North is up and east is to the left. {\it Right:} A zoomed image  on the cluster inner $3\arcmin\times3\arcmin$. Overlaid here are the surface mass density (white, starting at $3\sigma_\kappa$, in $2\sigma_\kappa$ increments) and X-ray brightness contours (red,  $2.5-9\sigma_\mathrm{X}$ in logarithmic spacing) from {\it Chandra}, depicting the gas density. Notably, there is no WL mass overdensity at the location of the X-ray interloper, but there is one located eastern of it. Additionally, the second most massive western substructure detected in our WL map is not traced by the X-ray map.}
\label{fig:color}
 \end{center}
\end{figure*}
A previous joint SL+WL study of this cluster was presented in M11, combining {\it HST}/ACS data along with wide-field imaging from VLT/VIMOS and the Subaru Telescope. However, The Subaru observations were taken under poor conditions, and in only one band ($\mathrm{FWHM}_{i'}\sim1.5\arcsec$). Therefore, M11 lensing analysis was governed by the SL+WL in the inner $3\arcmin\times 3\arcmin$, whereas its wide-field WL analysis was shallow and limited to the central $9\arcmin\times 9\arcmin$. 

We analyze here new data obtained with the wide-field camera Suprime-Cam 
\citep[$30\arcmin\times 30\arcmin$][]{Miyazaki2002a}
at the prime focus of the 8.3-m Subaru on the nights of 2013 July 14 \& 15, taken
in the $B,\RC,z'$ bands. We add to these the old data from Subaru $i'$ band, archival ESO/WFI\footnote{\url{http://archive.eso.org}} $V$ (ESO-843; program 079.A-0805) band, and archival VLT/HAWK-I $K_{\mathrm S}$ band (program 092.A-0472), in order to improve the accuracy of the photometric redshifts, used to determine the sample depth (\autoref{subsec:depth}).
The observation details of A2744 are summarized in \autoref{tab:data}.
\autoref{fig:color} shows a $BR_{\mathrm C}z'$ composite color image of the cluster central $26\,\arcmin\times26\,\arcmin$ (left panel), produced  with the {\sc Stiff} software \citep{Bertin2012}. We overlay it with the total mass density map determined from WL (white contours), the luminosity density map (yellow contours on the left panel) and the smoothed {\it Chandra} X-ray luminosity map (red contours on the right panel) which we describe in \autoref{subsec:2DK}.

Our reduction pipeline derives from \cite{Nonino+2009} and SDFRED \citep{Yagi+2002_SDFRED,Ouchi+2004_SDFRED} and has been
optimized for accurate photometry and WL shape measurements. It has been fully described in \cite{Medezinski2013}. We refer the reader to this work for full details and briefly summarize the main steps here. Standard reduction steps include bias subtraction, flat and super-flat field correction and
point-spread function (PSF) matching between exposures in the same band (to some intermediate level). Masking of saturated star trails and other artifacts is then applied.
We derive an astrometric solution with the SCAMP
software \citep{Bertin+2006_SCAMP} using VISTA/VIRCAM z'-band image from the ESO archive, which in turn has been tied to 2MASS\footnote{This publication makes use of data products from the Two Micron All Sky Survey, which is a joint project of the University of Massachusetts and the Infrared Processing and Analysis Center/California
Institute of Technology, funded by the National Aeronautics and Space
Administration and the National Science Foundation} as an external reference
catalog. 
Finally, the {\sc Swarp} software \citep{Bertin2002} is utilized to stack the single exposures on a common WCS grid with pixel-scale of $0.2\arcsec$.

The photometric zero-points were
derived from a suitable set of Standard stars. These zero-points were refined by fitting spectral energy distribution (SED) templates with the BPZ code \citep[Bayesian photometric redshift
estimation,][]{BPZ,Benitez+2004} to the colors of $1311$ galaxies
having spectroscopic redshifts  from \cite{Owers2011}. This leads to a final photometric accuracy of $\sim 0.005-0.02$\,mag in the different passbands.
The 6-band $BVR_{\mathrm C}i'z'K_{\mathrm S}$ photometry catalog was measured
using SExtractor \citep{Bertin2002} in dual-image mode on
PSF-matched images created by ColorPro
\citep{coe06}, where a combination of $B+V+R_{\mathrm C}+i'+z'$ bands was used as a
deep detection image. The stellar PSFs were measured from 100 stars and modeled using {\sc IRAF} routines. 
The number density of galaxies in our photometric catalog is $\bar{n}_g\sim64$\,arcmin$^{-2}$.

\subsection{Shape Measurement}
\label{subsec:shape}

For shape measurements, we use our
WL analysis pipeline which is based on the IMCAT package \citep[KSB
hereafter]{KSB}, incorporating modifications and
improvements developed and outlined in \citet{Umetsu2010}.
Our KSB+ implementation has been applied extensively to
Subaru/Suprime-Cam cluster observations
\citep[e.g.,][]{Coe2012,Merten2015,Medezinski2010,
Medezinski2011,Medezinski2013,Okabe+Umetsu2008,UB2008,
Umetsu+2009,Umetsu2010,Umetsu+2011,Umetsu+2011_stack,Umetsu+2012,Zitrin+2011_A383,Zitrin2013}. 
Full details of our WL analysis pipeline are presented in
\cite{Umetsu+2012} and \cite{Umetsu2014}. Here we reiterate some of the key aspects of the pipeline for completeness.
Objects are detected using the {\sc IMCAT}  peak finding algorithm, {\sc hfindpeaks}, providing a peak position, Gaussian scale length, $r_g$, and an estimate of the significance of the peak detection, $\nu$.  Close pairs of objects detected above  $\nu=7$ and having a neighbor within $3r_g$ are rejected to avoid shape measurement bias. Then all objects with  $\nu<10$ are excluded from the analysis.  KSB's isotropic correction factor, $P_g$, is calibrated as a function of object size ($r_g$) and magnitude,  
using galaxies detected with high significance $\nu>30$
\citep{Umetsu2010}, so as to minimize the inherent shear calibration bias. Finally, we apply it to obtain the shape estimate as $g_\alpha=e_\alpha/P_g$, with $e_\alpha$ being the      anisotropy-corrected ellipticity.

In this analysis we use the $\RC$-band for shape measurement, which was taken in medium seeing 
conditions, but is significantly better compared to the earlier
$i'$-band data taken in 2008.  
We do not PSF-match the single exposures before stacking (as done for
photometric measurements), to retain the WL
information derived from the shapes of galaxies. 
In addition to the $\nu$ cuts described above,
we apply the following stringent size cuts \citep{Umetsu+2012}
when deriving the final shape catalog:
$r_g > \mathrm{mode}(r_g^*)$ and 
$r_h > \mathrm{mode}(r_h^*) + 1.5\sigma(r_h^*)$,
where $r_h$ is the object half-light radius  and the subscript asterisk
denotes quantities for stellar objects.

We restrict the catalog to the central $26\arcmin\times26\arcmin$ region
to ensure accurate PSF characterization.
%
Based on simulated Subaru Suprime-Cam images \citep[see Section 3.2 of][]{Oguri2012,Massey2007},
we include in our analysis a calibration factor of $1/0.95$ as
$g_i\to g_i/0.95$ to account for residual calibration.
The resulting shape catalog has a number density of galaxies, $\bar{n}_g\sim25$\,arcmin$^{-2}$, including cluster members and foreground galaxies.

\subsection{Sample Selection}
\label{subsec:samples}


\begin{deluxetable}{ccccccc}
\tabletypesize{\footnotesize}
\tablecolumns{7} 
\tablecaption{ \label{tab:samples} Galaxy color selection.}  
\tablewidth{0.5\textwidth}  
\tablehead{ 
 \multicolumn{1}{c}{Sample} &
 \multicolumn{1}{c}{Magnitude limits} &
 \multicolumn{1}{c}{$N_g$} &
 \multicolumn{1}{c}{$\bar{n}_g$} & 
 \multicolumn{1}{c}{$\langle z_\mathrm{s}\rangle$} &
 \multicolumn{1}{c}{$z_{{\mathrm s, eff}}$} &
 \multicolumn{1}{c}{$\langle \beta \rangle$} 
\\
 \colhead{} & 
 \multicolumn{1}{c}{(AB mag)} &
 \colhead{} &
 \multicolumn{1}{c}{($\arcmin^{-2}$)} &
 \colhead{} &
 \colhead{} &
 \colhead{} 
} 
\startdata  

green & $16.3<z'<23.5$ & 1135 & 1.1 & 0.3 & 0.3 & $0.07\pm0.07$  \\ 
red & $21.0<z'<25.6$ & 4149 & 5.7 & 1.4 & 1.2 & $0.68\pm0.07$  \\ 
blue & $22.0<z'<26.7$ & 3777 & 5.1 & 1.7 & 1.2 & $0.67\pm0.16$  \\ 
red+blue & $21.0<z'<26.7$ & 7926 & 10.8 & 1.5 & 1.2 & $0.68\pm0.11$  \\ 

\enddata 
\tablecomments{
Column~(1): samples used in this analysis; the red and blue samples statistics are derived after matching with the shape catalog.
Column~(2): Magnitude limits for the galaxy sample. 
Column~(3): Number of galaxies.
Column~(4): Mean surface number density of galaxies.
Column~(5): Mean photometric redshift of the sample obtained with the BPZ code.
Column~(6): Effective source redshift corresponding to the mean depth $\langle\beta\rangle$ of the sample.
Column~(7): Distance ratio averaged over the redshift distribution of the sample.}
\end{deluxetable}


Here we describe our selection of cluster and
background galaxy samples, trying to maximize the number of usable galaxies for our WL analysis while avoiding dilution of the WL signal by cluster and foreground members.  We
use the $B,R_{\mathrm C},z'$ Subaru imaging which spans the full optical wavelength range to perform color-color (CC) selection, as established in \cite{Medezinski2010}.

\subsubsection{Cluster Sample}
\label{subsec:clsample}
We plot the $B-\RC$ vs $\RC-z'$ distribution of all galaxies from our photometric sample to the limiting magnitude (\autoref{fig:CC}, cyan).
We  identify the cluster-dominated area in this CC-space by separately plotting the number density in $B-R_{\mathrm C}$ vs.
$\RC-z'$ using only galaxies
having small projected cluster-centric radii, $\theta< 2\arcmin$ ($\simlt 0.4$\,Mpc at
$z_\mathrm{l}=0.308$). We specify a region above some characteristic overdensity in this space (shown as a solid green curve in \autoref{fig:CC}).
Then, {\bf all} galaxies within this distinctive region from the full CC
diagram define the ``green'' sample (\autoref{fig:CC}, green points),
comprising mostly the red-sequence of the cluster and a blue trail of
later-type cluster members.
This is also demonstrated by overlaying spectroscopically-selected cluster members  (\autoref{fig:CC}, black points), which nicely coincides with the location of the green sample where the red-sequence lies.

The WL signal for the ``green'' population is consistent with zero at all radii (\autoref{fig:gt1D}, green crosses), confirming it contains no background.
For this population of galaxies, we find a mean photometric redshift of
$\langle z_{\mathrm phot}\rangle \approx 0.305$ (see \autoref{subsec:depth}), consistent with the cluster redshift.
Importantly, the green sample marks a region that contains a majority
of unlensed galaxies, relative to which we select our background
samples, as summarized below.

\subsubsection{Background Sample}
\label{subsec:bgsample}

\begin{figure}[tb]
 \begin{center}
\includegraphics[width=0.48\textwidth,clip]{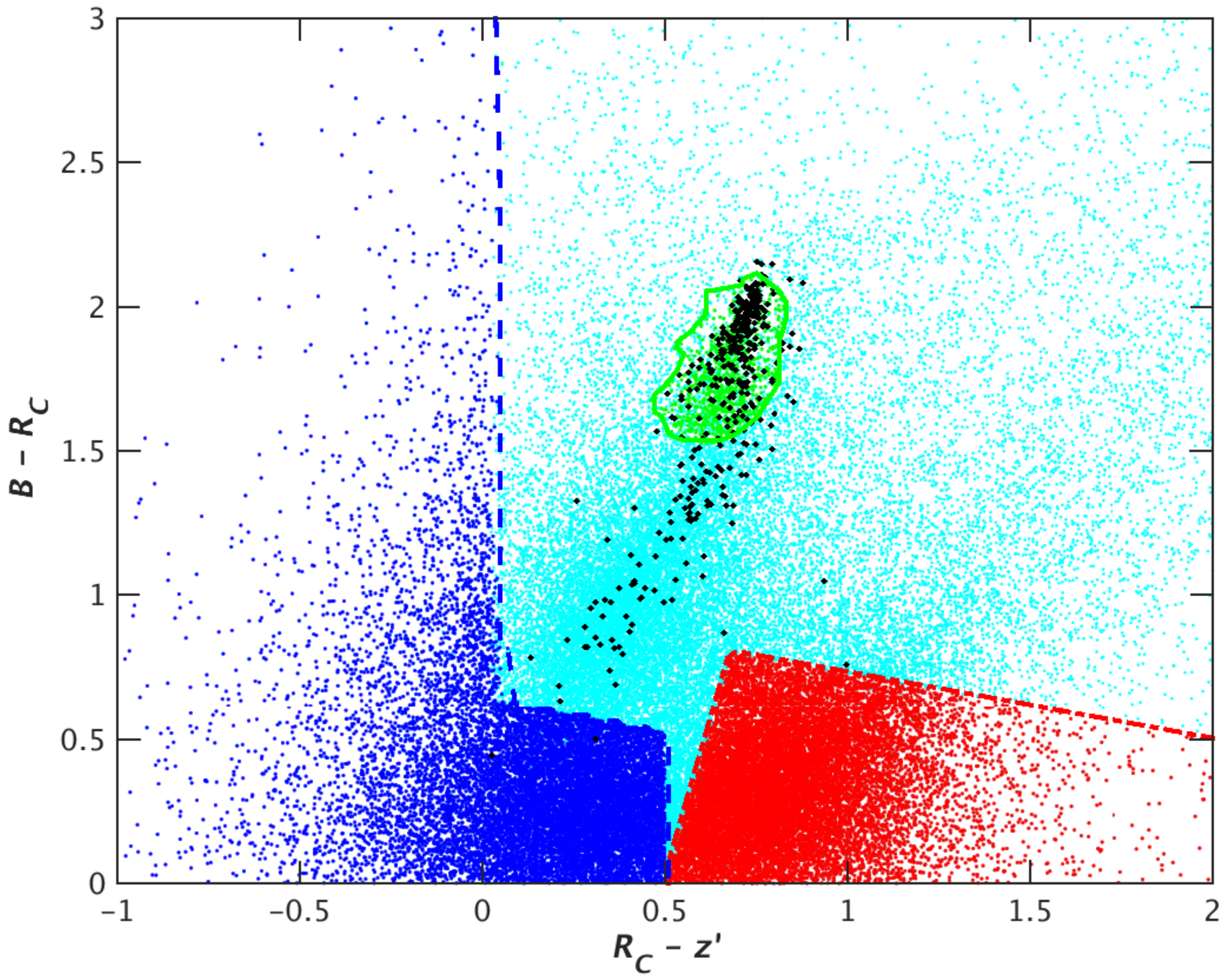}
 \end{center}
\caption{ $B-\RC$ vs. $\RC-z'$ color-color diagram of all the galaxies
  in the Subaru imaging (cyan).  ``Blue'' and ``red'' background
  galaxies (lower-left blue and lower-right red, respectively) are
  selected for WL mass analysis. The ``green'' sample (green solid
  contour), defined by an overdense region of small clustercentric
  radii galaxies, comprises mostly red sequence and some blue later
  type cluster members.  The background samples are well isolated from
  the green region and satisfy other criteria as discussed in
  \autoref{subsec:bgsample}. Our background selection successfully
  excludes all spectroscopically confirmed cluster members (black; though see text for details).}
\label{fig:CC}
\end{figure}

It is critical to avoid inclusion of unlensed cluster members and foreground galaxies in the background sample used for the WL analysis, so as not to dilute
the true lensing signal 
\citep{Broadhurst05b,Medezinski2007,Medezinski2010}.
This dilution effect reduces the strength of the lensing
signal \citep[by a
factor of 2--5 at $R\simlt 400\,{\mathrm kpc}\,h^{-1}$; see 
Figure 1 of][]{Broadhurst05b}, particularly at small radii
where the cluster potential is largest, in proportion to the fraction of
unlensed galaxies whose orientations are randomly distributed. 

We use the CC selection method of \citet{Medezinski2010} to
define undiluted samples of background galaxies,
\citep[for details, see][]{Medezinski2010,Umetsu2010}, which has been successfully applied to many cluster WL analyses \citep{Medezinski2010,Medezinski2011,Medezinski2013,Merten2015,Umetsu2010,Umetsu+2011,Umetsu+2012,Umetsu2014,Umetsu2015}.
We make use of the $(B-R_{\mathrm C})$ vs. $(R_{\mathrm C}-z')$
CC-diagram to encompass the red and blue branches of galaxies.
of the WL signal is visible, to minimize contamination by unlensed
cluster members and foreground galaxies.  The color boundaries of our
``blue'' and ``red'' background samples are shown in
\autoref{fig:CC}. The magnitude limits of all the samples are given in
\autoref{tab:samples}.  For both the blue and red samples, we find a
consistent tangential shear profile (see \autoref{subsec:gt}), rising
all the way to the cluster center, as expected (see
\autoref{fig:gt1D}).  Furthermore, none of the
spectroscopically-selected cluster members (black points in
\autoref{fig:CC}) are present in the blue or red samples, confirming
the purity of our sample. The two black points that
only appear to be within the color boundaries of the blue background
sample are in fact not included in it, as the faintest spec-z member
($z'=21.8$) is brighter than the bright blue magnitude limit
($z'>22$). Finally, The percentage of galaxies in the background
sample that have photo-$z$'s (see \autoref{subsec:depth}) smaller than the cluster redshift (all
of which are in the blue sample) is small, just 1.6\% of the
background sample, demonstrating a negligible contamination level.

Full details of our samples are listed in \autoref{tab:samples}.  Overall, our
CC-selection criteria provides $N_g= 1135,17141,$ and 16281
galaxies, in the green, red, and blue photometry samples,
respectively.  For our WL distortion analysis, we have a subset of
4149 and 3777 galaxies in the red and blue samples (with usable $\RC$
shape measurements), respectively, or a total background number
density of $\bar{n}_g=10.8$\,arcmin$^{-2}$.

\subsection{Depth Estimation}
\label{subsec:depth}

\begin{figure}[tb]
 \centering
\includegraphics[width=0.48\textwidth]{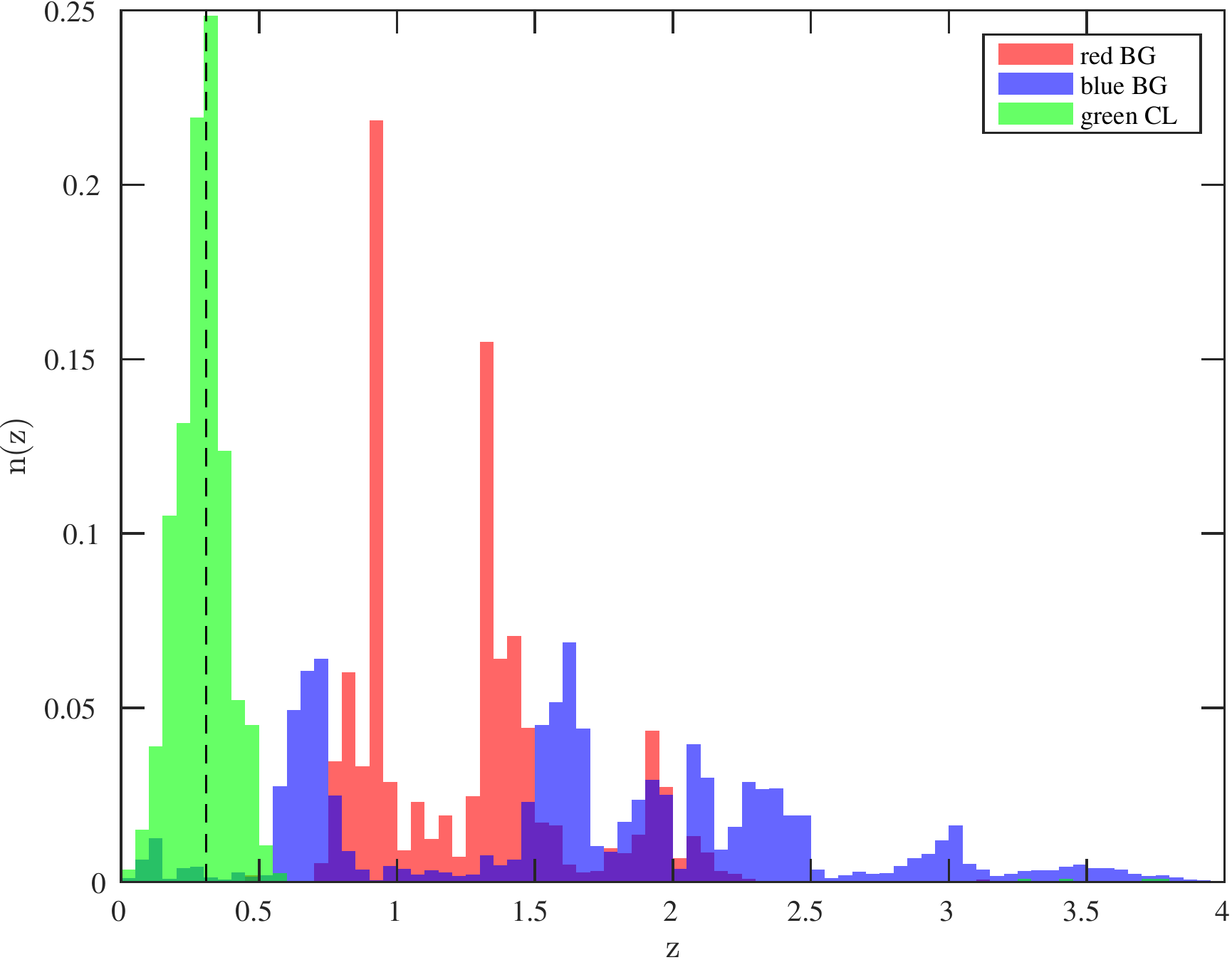}
\caption{Normalized redshift distributions of green, red and blue samples defined by the CC-selection. Redshift estimates are derived from BPZ photo-$z$'s. The cluster redshift is marked with a black line. }
\label{fig:zdist}
\end{figure}

We use the {\sc BPZ} code to measure photometric
redshifts (photo-$z$s) $z_{\mathrm phot}$ from our  
deep Subaru+WFI+HAWK-I $BVR_{\mathrm C}i'z'K_{\mathrm s}$ photometry (\autoref{subsec:data}, though note $K_{\mathrm s}$ only covers the central $4\arcmin$). We present the redshift distributions of the red, blue and green samples in \autoref{fig:zdist}. As can be seen, the green sample lies at about the cluster redshift, $z_{\mathrm l}=0.308$, whereas the red and the blue samples lie mostly at higher redshifts, $z_{\mathrm phot}\simgt0.5$. According to the photo-$z$'s of our background sample (\autoref{subsec:bgsample}), the level of foreground contamination is negligible ($<2\%$). We emphasize, that although the redshift distribution of our CC-selected sample is good enough to demonstrate the reliability of our selection of background-only galaxies, the reverse is not true -- using photo-$z$'s for background selection that are based on only $\sim5$ optical bands is not sufficiently unbiased, as photo-$z$'s are highly degenerate, especially for blue galaxies that have relatively flat SEDs. We, on the other hand, using CC-selection, can select those faint blue background galaxies that lie at high redshifts, $z_\mathrm{phot}=2-3$  (bottom-left corner of \autoref{fig:CC}).
 
To convert to physical mass units, we need to estimate the mean depths ($\langle\beta\rangle$, see \autoref{sec:theory}) of the background samples used in our WL analysis. We follow the prescription devised in \cite{Umetsu2014}, and exclude galaxies above $z_\mathrm{phot}>2.5$ and having ODDS$<0.8$ (as given by the BPZ code). In \autoref{tab:samples} we summarize the mean depths 
$\langle\beta \rangle$ and the effective source redshifts 
$z_\mathrm{s,eff}$ for our background samples.
The mean depth for the combined blue and red sample of background galaxies is $\langle\beta_\mathrm{back}\rangle=0.68\pm 0.11$, which corresponds to $z_\mathrm{ s,eff}=1.2\pm 0.1$.

\section{WL Analysis}
\label{sec:WL}

\subsection{Tangential Distortion Analysis}
\label{subsec:gt}

The tangential component of the
reduced-shear, $g_{+}$, is used to obtain the azimuthally-averaged
distortion due to lensing:
\begin{equation}
g_{+}=-( g_{1}\cos2\theta +  g_{2}\sin2\theta),
\end{equation}
where $\theta$ is the position angle of an object with respect to the
cluster center, and $g_{1},g_{2}$ are the Cartesian distortion coefficients.

\begin{figure}[tb]
\includegraphics[width=0.48\textwidth]{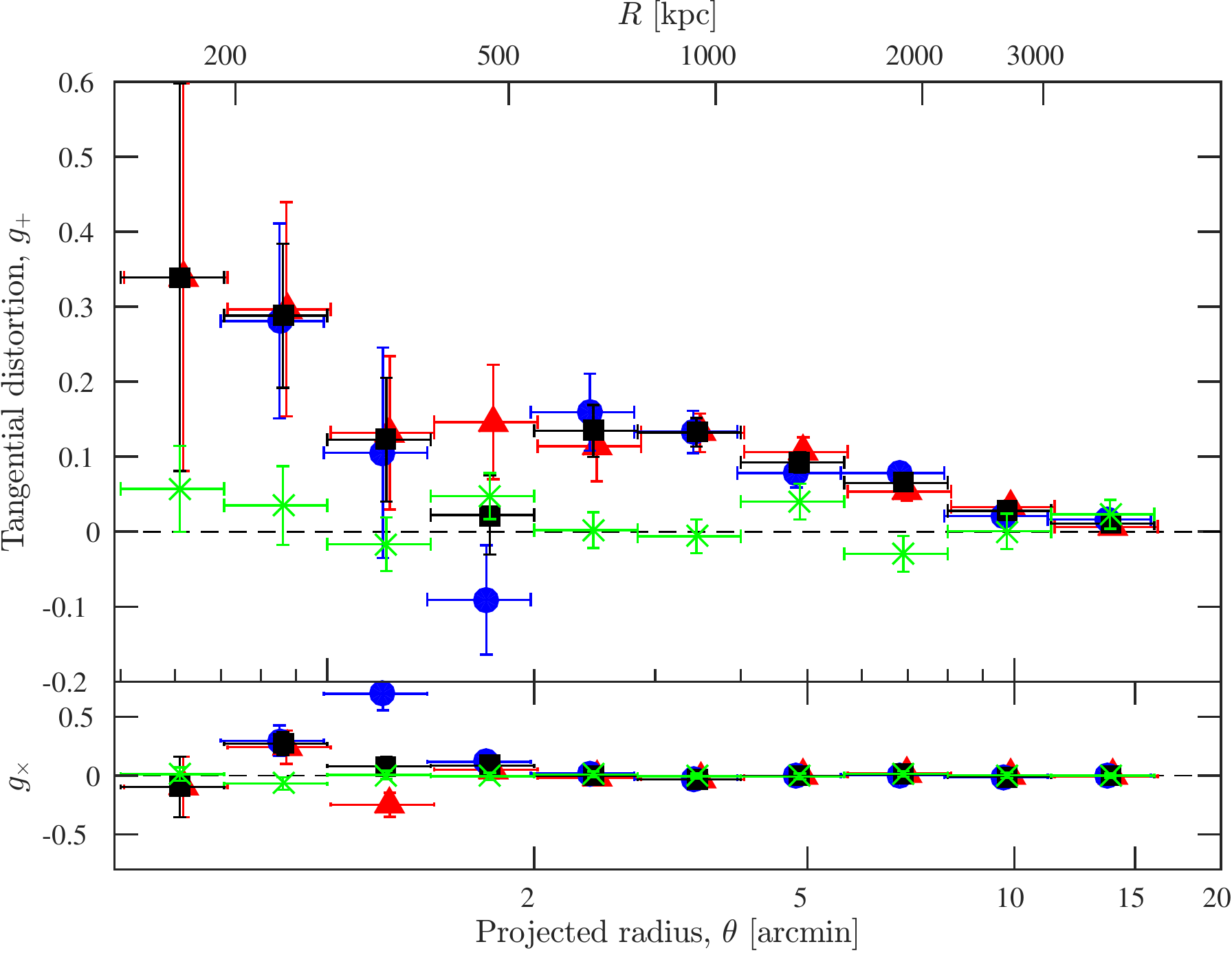}
\caption{ Binned tangential reduced-shear $g_+$ (upper panel) and the
  $45^\circ$ rotated ($\times$) component $g_{\times}$ (lower panel)
  as a function of cluster radius, for our red (triangles),
  blue (circles), green (crosses) and blue+red (squares) galaxy
  samples, horizontally shifted for visual clarity. The
  $\times$-component is consistent with a null signal detection to
  within $1\sigma$, except in the strong regime,
  $\theta\simlt2\arcmin$, where the cluster morphology is complicated
  by substructure.}
\label{fig:gt1D}
\end{figure}

In \autoref{fig:gt1D} we plot the $g_+$ radial profile for
the green, red and blue samples. The black points represent the
red+blue combined sample.  The shear profiles obtained from both the
red and blue sample rise toward the center of the cluster, and agree
with each other within the errors, demonstrating that both these
samples are dominated by background galaxies. It is only below
$\theta\simlt2\arcmin$ that we find some disagreement, likely due to the inner cluster
substructure.  The $g_+$ profile of the
green sample agrees with zero at all radii, supporting the
interpretation that it consists of mostly cluster members.

\subsection{Total Mass, Gas and Light Distributions}
\label{subsec:2DK}
\begin{figure*}[htb]
\centering
\includegraphics[width=0.49\textwidth]{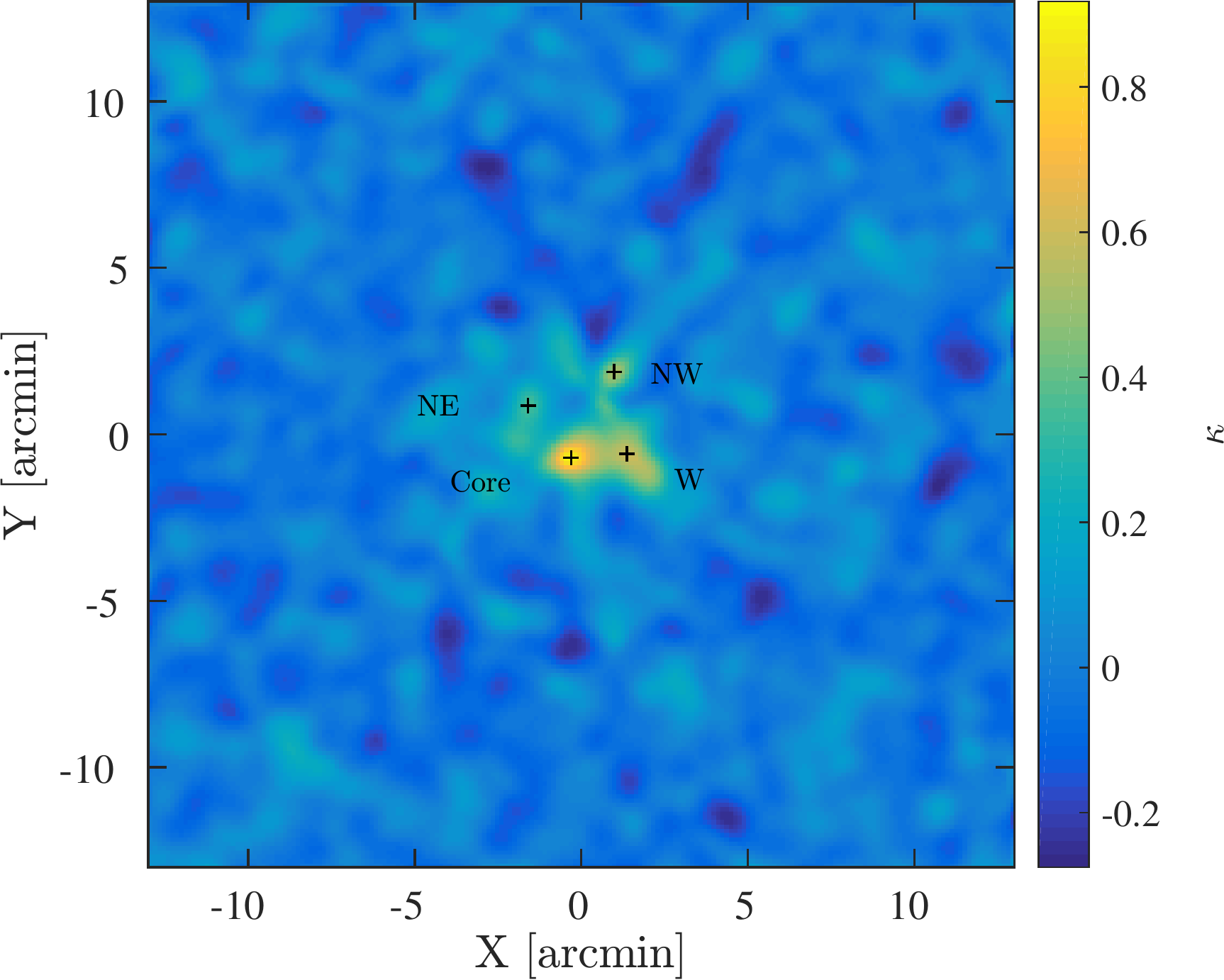}
\includegraphics[width=0.48\textwidth]{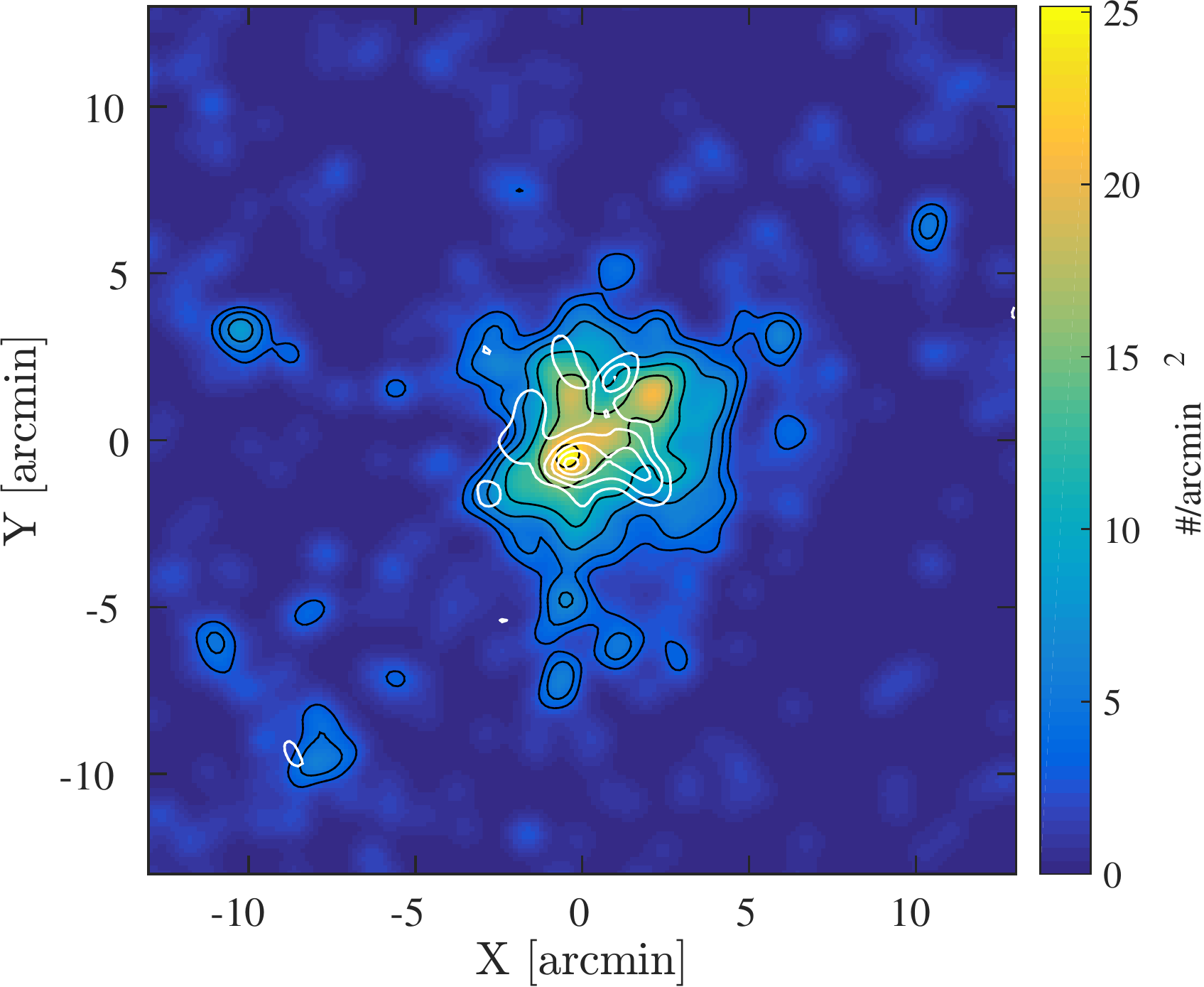}
\includegraphics[width=0.48\textwidth]{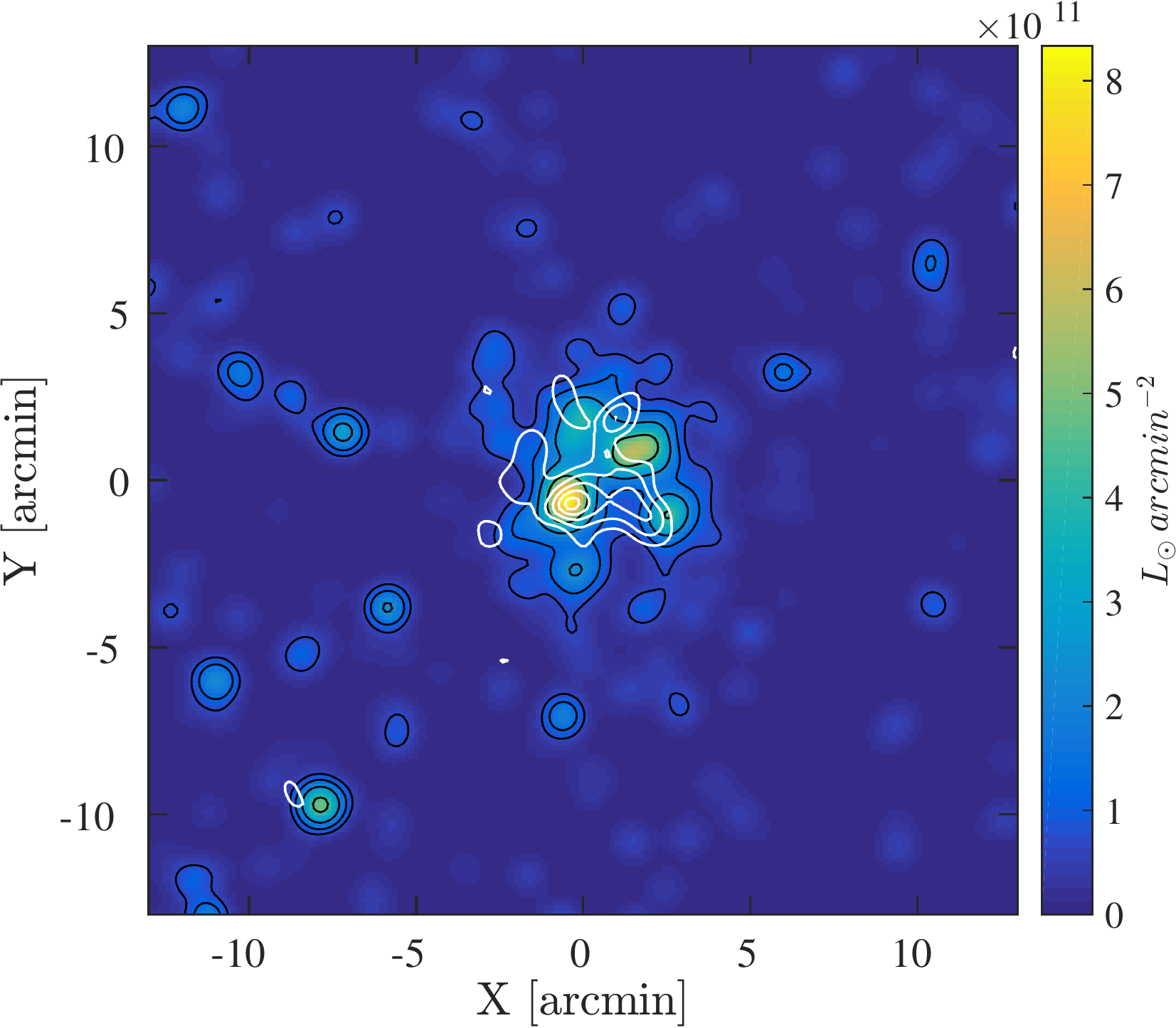}
\includegraphics[width=0.5\textwidth]{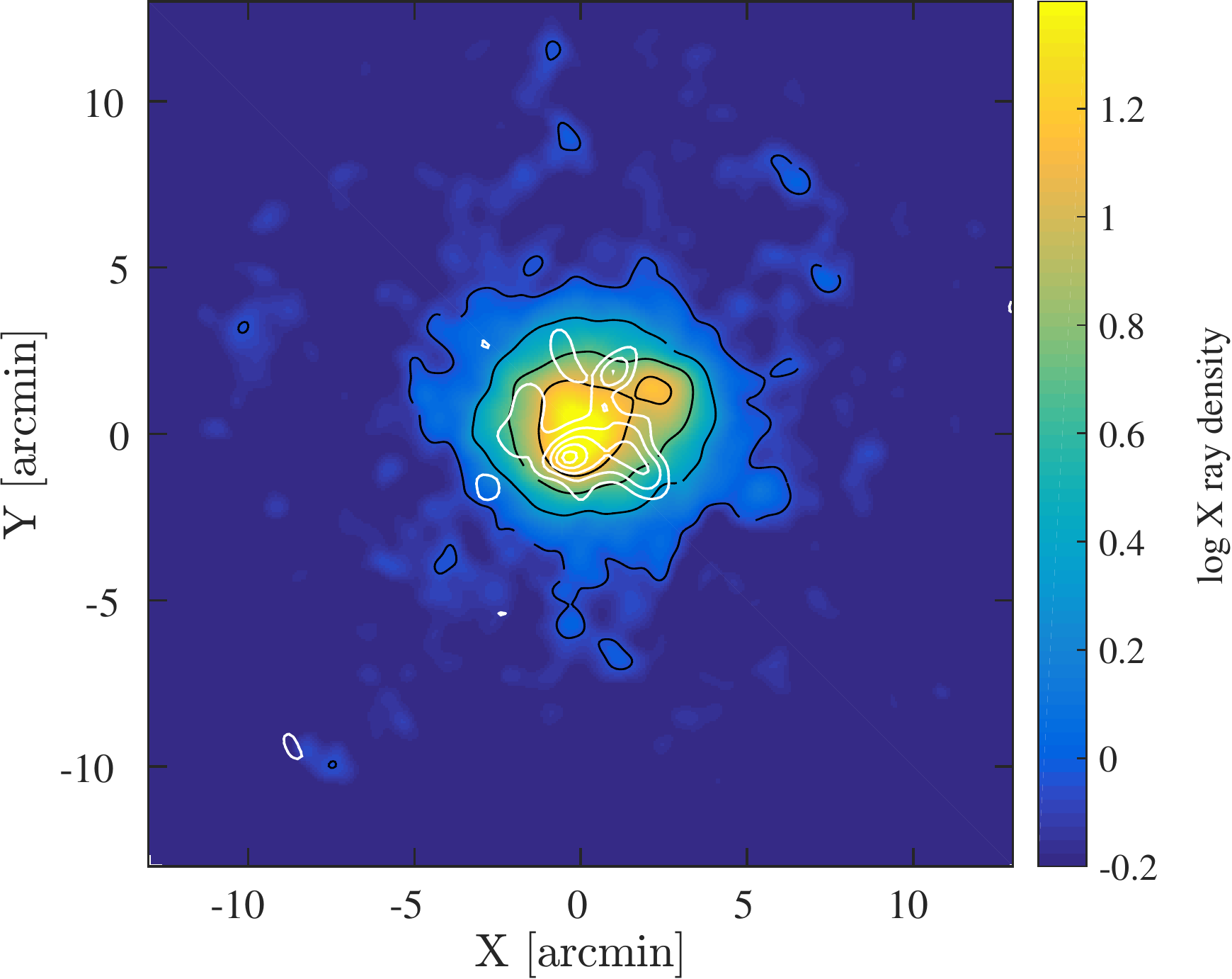}

\caption{{\it Top Left}: The dimensionless surface-mass density
distribution, or the lensing convergence $\kappa(\btheta)$, reconstructed from Subaru distortion data. Black pluses denote the four density substructures comprising the cluster, detected at a significance level $>4.5\sigma_\kappa$. 
{\it Top Right}: Surface number density distribution 
of cluster member galaxies. Also overlaid are the white contours of the surface-mass density map above $3\sigma_\kappa$, in  $2\sigma_\kappa$ increments.
{\it Bottom Left}: Luminosity density distribution 
of cluster members, overlaid with the same surface-mass density contours (white).
{\it Bottom Right}: X-ray brightness distribution from {\it XMM-Newton}, overlaid with the same surface-mass density contours (white).
}
\label{fig:kappa2D}
\end{figure*}
To recover the underlying projected mass density field
$\Sigma(\btheta)$, we use the \cite{KaiserSquires93} linear inversion method
 \citep[also see][Section 4.4]{Umetsu+2009}.
We first pixelize the distortion measurements of the background
(blue+red) sample (\autoref{subsec:bgsample}) using \autoref{eq:ggrid}
onto a $200\times200$ regular grid over the central 
$26\arcmin\times 26\arcmin$ region, with a Gaussian kernel,
$S(\theta)\propto \exp [-\theta^2/2\theta_\mathrm{g}^2]$, where
$\theta_{\mathrm g}={\mathrm FWHM}/\sqrt{8\ln{2}}$ is the smoothing scale. We set $\mathrm{FWHM}=1\arcmin$,
chosen to optimally balance the number density of background galaxies
available for the WL reconstruction, and to resolve the adjacent
substructures in the core of the cluster, separated by an order of
$\sim2\arcmin$.  We then invert the distortion map using \autoref{eq:KS93} to obtain the lensing convergence field, 
$\kappa(\btheta)$, shown in \autoref{fig:kappa2D} (top left
panel). 

We compare the mass density map with both a 2D galaxy number-density map (\autoref{fig:kappa2D}, top right panel) and a $K$-corrected $\RC$-band luminosity density map (\autoref{fig:kappa2D}, bottom left panel) of the cluster members (utilizing the green sample). Both maps are smoothed with a Gaussian kernel of the same scale as the mass map above. We also overlay the surface mass density map starting at $3\sigma_\kappa$, in $2\sigma_\kappa$ increments (white contours), to illustrate the correlation between these. 

Finally, we utilize both {\it Chandra} and {\it XMM-Newton} public X-ray observations to examine how the gas is distributed in the core and on larger scales. We acquire public {\it XMM-Newton} images from the XSA archive (PI Bohringer; totaling 11,587 sec). We perform point-source removal and interpolation on the MOS1+MOS2+pn stacked image, and smooth the image to FWHM$=0.667\arcmin$. We present the map of X-ray emission on large scales from the {\it XMM} observations in \autoref{fig:kappa2D} (bottom right), along with the mass density contours (white).
{\it Chandra} provides high-resolution imaging of the gas in the core of the cluster ($3\arcmin\times3\arcmin$). We overlay the X-ray brightness contours from the {\it Chandra} stacked image using the dataset as processed by M11 in \autoref{fig:color} (right panel, red contours), and also in \autoref{fig:BS}, starting at a higher level in order to demonstrate the fine details of the gas substructure. The main diffuse gas cloud is seen to lie between the cluster galaxy components, whereas an X-ray interloper is detected in the NW direction but beyond the BCGs in that location. The cool core ($T_\mathrm{X}=7.7$\,keV) discussed in \cite{Owers2011} is detected in the bottom (south) of the main gas cloud. We will further discuss the correlations between the gas, stars and DM in \autoref{subsec:Msep}, but first we need to model the different cluster DM components, which we present in \autoref{subsec:2Dhalo}.


\subsection{Substructure Detection}
\label{subsec:substr}


\begin{deluxetable}{lrrccc}
\tabletypesize{\footnotesize}
\tablecolumns{6} 
\tablecaption{ \label{tab:subhalos} Detected Substructures.}  
\tablewidth{0.5\textwidth}  
\tablehead{ 
 \multicolumn{1}{l}{Halo} &
 \multicolumn{1}{c}{$X_\mathrm{c}$} &
 \multicolumn{1}{c}{$Y_\mathrm{c}$} &
 \multicolumn{1}{c}{$z_l$} & 
 \multicolumn{1}{c}{$z_{{\mathrm s, eff}}$} &
 \multicolumn{1}{c}{$S/N$} 
\\
 \colhead{} & 
 \colhead{(arcmin)} &
 \colhead{(arcmin)} &
 \colhead{} &
 \colhead{} &
 \colhead{} 
} 
\startdata  

Core& $-0.33\pm0.18$ & $-0.72\pm0.16$& $0.32\pm0.01$ & $1.27\pm0.08$ & 12.1 \\
W   & $1.36\pm0.42$ & $-0.59\pm0.43$  & $0.30\pm0.04$ & $1.27\pm0.07$ & 7.9 \\
NE  & $-1.62\pm0.33$ & $0.85\pm0.53$& $0.30\pm0.02$ & $1.27\pm0.08$ & 4.7 \\
NW  & $0.98\pm0.37$ & $1.89\pm0.48$ & $0.30\pm0.01$ & $1.27\pm0.08$ & 7.0 \\

\enddata 
\tablecomments{
Column~(2--3): The centroid positions of each halo, given relative to the cluster center in units of arcminutes. Errors are derived as the centroid standard deviation in the bootstrapped maps.
Column~(4): lens redshift, determined from ``green'' galaxies within $1\arcmin$ of the peak location.
Column~(5): effective source redshift, estimated from the background sample.
Column~(6): Signal-to-noise ratio determined from the convergence map.}
\end{deluxetable}

\begin{figure}
 \centering
\includegraphics[width=0.48\textwidth]{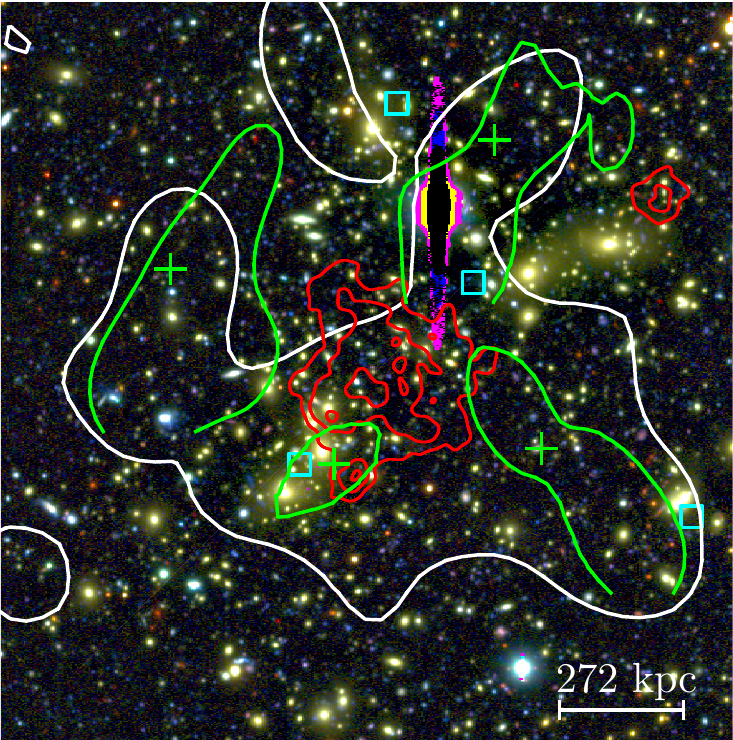}
\caption{As in \autoref{fig:color}, color image of the cluster center ($3\arcmin\times3\arcmin$), showing the $3\sigma_\kappa$ surface mass density contour (white), X-ray brightness contours (red, in exponential scale in order to enhance substructure) from {\it Chandra}, and the $90\%$ confidence bounds on the centers of the detected substructures, as derived from 500 bootstrapped convergence maps. The four substructures detected in the full constrained convergence map are marked by green pluses. 
We also mark the location of the four peaks found by M11 (cyan squares).}
\label{fig:BS}
\end{figure} 

We are able to detect substructures in our convergence map down to spatial separations of $\sim1\arcmin$, which is the smoothing scale of our map. We apply a detection algorithm on our mass map, finding local maxima above $4.5\sigma_\kappa$, with separations of $>1\arcmin$. We find four individual peaks with significance $>4.5\sigma_\kappa$. These are labeled as Core, W, NE, NW in the convergence map (\autoref{fig:kappa2D}, top left, black pluses), detected at significance levels of $12.1\sigma_\kappa$, $7.9\sigma_\kappa$, $4.7\sigma_\kappa$ and  $7.0\sigma_\kappa$, respectively. These are also labeled in \autoref{fig:color} (right panel), and marked in \autoref{fig:BS} (green pluses). We summarize the details of the detected subhalos in \autoref{tab:subhalos}.

As discussed in \autoref{subsec:samples}, WL analyses may be subject to
several sources of systematics. Two main sources are the contamination
of the background sample by foreground galaxies and the systematic
errors from galaxy shape measurements. As presented, we have attempted to address the first by a careful selection of the source sample according to the CC-selection method (see \autoref{subsec:bgsample}). As for the latter, we have excluded from our shape catalog noisy measurements using several cuts (see \autoref{subsec:shape}). Here we investigate how these two different systematics affect the resulting mass reconstruction map, and subsequently, the detected substructures.

\begin{figure*}
 \centering
\includegraphics[width=0.48\textwidth]{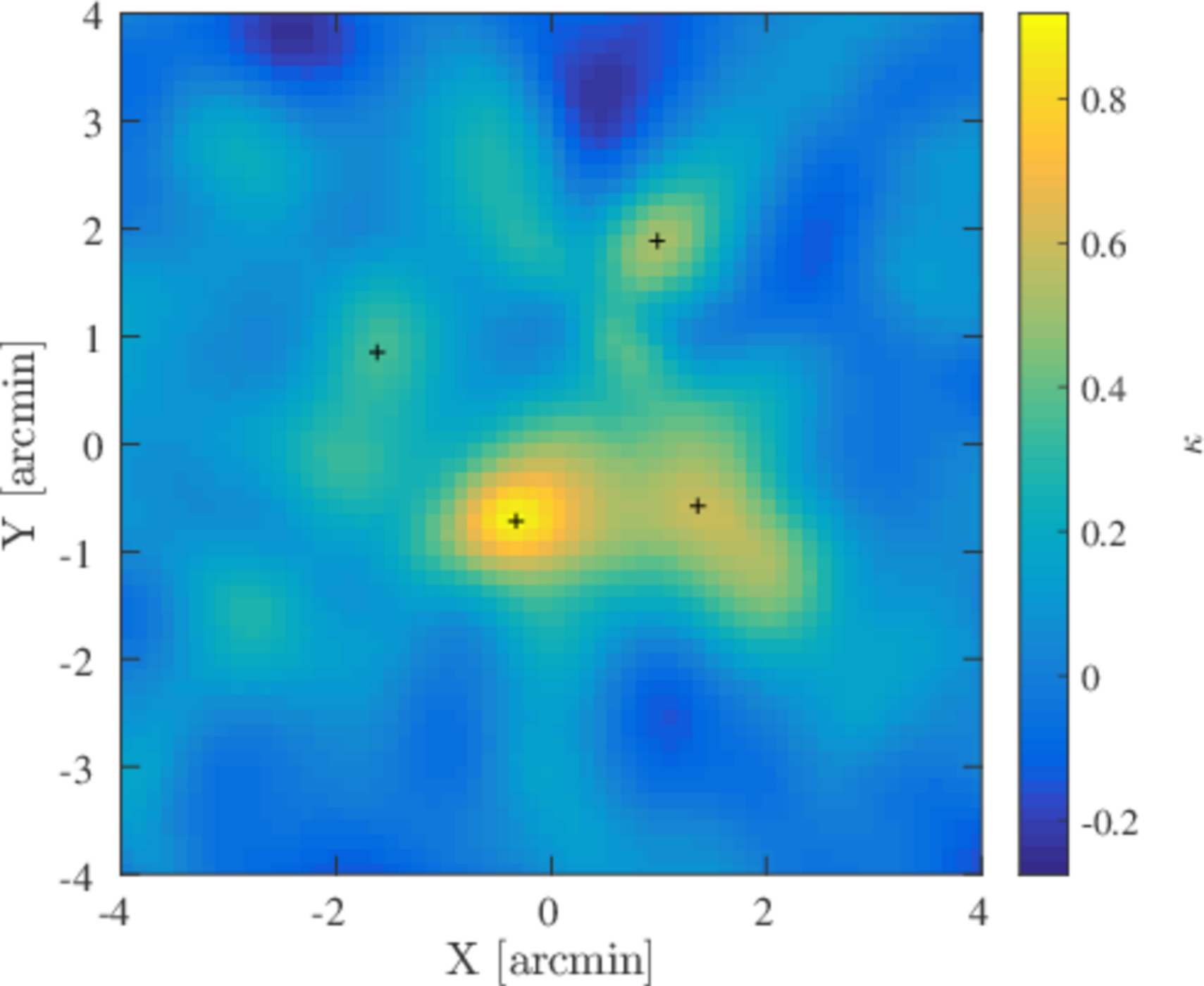}
\includegraphics[width=0.48\textwidth]{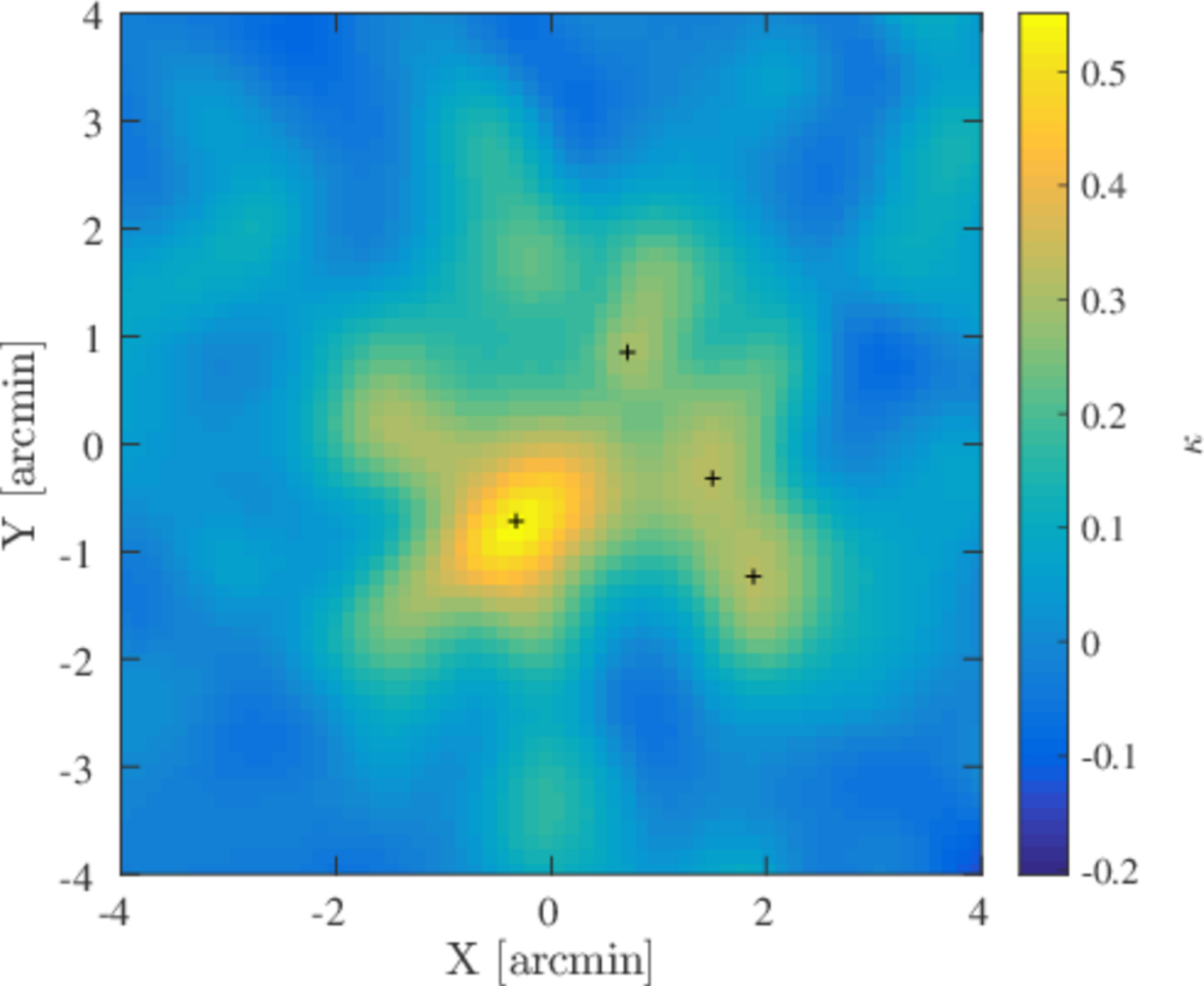}
\includegraphics[width=0.48\textwidth]{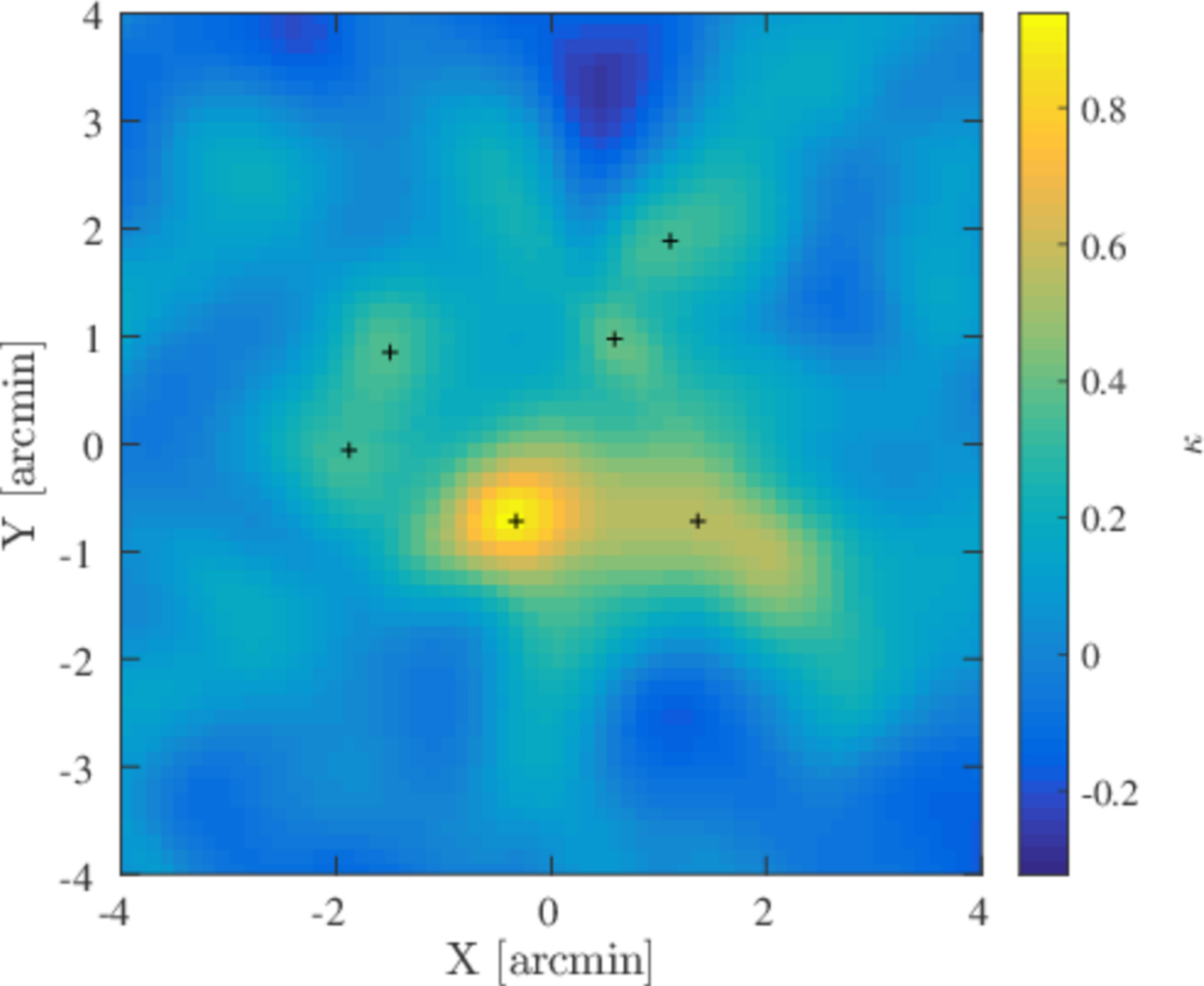}
\includegraphics[width=0.48\textwidth]{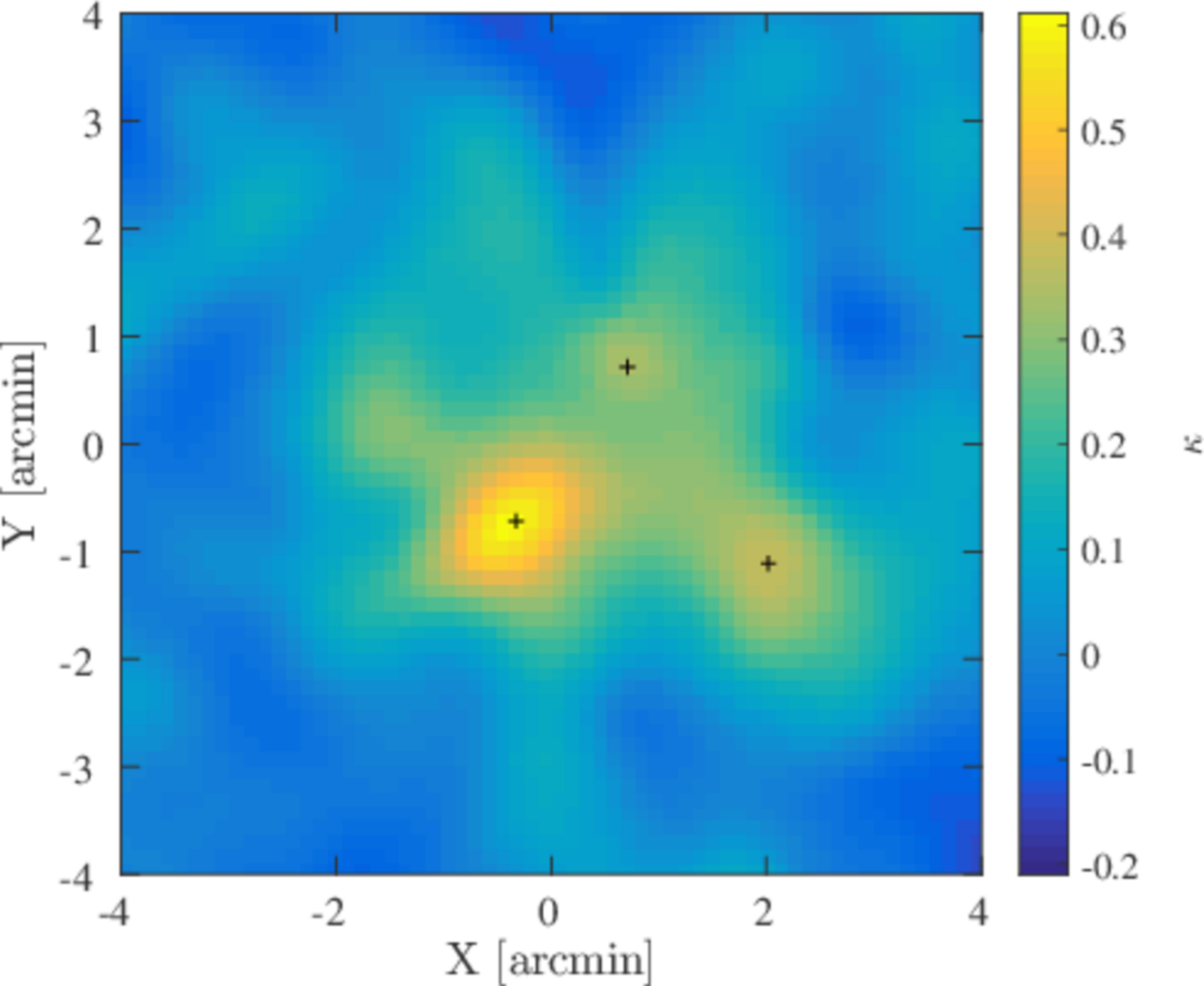}
\caption{Convergence maps for different shape catalogs and/or background selections. In each case, the black pluses mark the substructure peaks detected at above $>4.5\sigma_\kappa$. {\it Top left:} using the constrained shape catalog with CC selection applied, the same as in our analysis.
{\it Top right:} using the constrained shape catalog with photo-z selection applied.
{\it Bottom left:} using the biased shape catalog with CC selection applied.
{\it Bottom right:} using the biased shape catalog with photo-z selection, the most like M11's WL analysis.}
\label{fig:WLsys}
\end{figure*}

Following M11, we construct a Subaru shape catalog where we apply only the
$r_h>\mathrm{mode}(r_h^*)+\sigma_{r_h^*}$ cut, which we dub the {\em biased} shape catalog. Additionally, we perform a background selection based on photometric redshifts, requiring that $z_\mathrm{phot}>0.5$, as in M11. We explore all three combinations of these two variants, and construct a convergence map for each of the following: (1) 
our constrained shape catalog, but with photo-$z$ selection applied; (2) the biased shape catalog, but with our CC-selection applied; and (3) the biased shape catalog, with photo-$z$ selection applied. The latter is the most similar to the sample used in M11's analysis, and in most WL analyses in the literature. We display the results in \autoref{fig:WLsys}, where we compare with our original analysis in the top left panel, case (1) presented in the top right panel, case (2) in the bottom left panel, and case (3) in the bottom right panel.

Both these systematic effects change the resulting WL map significantly for the less massive substructures, whereas the main peak remains most significant and approximately in the same location. The most affected by these are the locations of the W and NW clumps, which appear closer to their location in M11's analysis in case (3) which is the closest to M11's WL selection. In both cases where the photo-$z$ selection is applied (cases (1) and (3)) the WL map suffers from significant dilution compared to the CC selection cases, making the core less significant.

Since background selection has such a large effect on the detected substructures' locations, we attempt to account for this systematic uncertainty by performing a bootstrap analysis using a wider source sample than our reference constrained CC-selected sample. For this, we join the above two samples, the photo-z selected (3) and CC-selected sample (2) from the biased shape catalog, together containing 18,068 galaxies, or a source density of $\bar{n}_\mathrm{g}=24.6$ arcmin$^{-2}$.
We draw from this joint sample 500 bootstrapped convergence maps, and apply the same detection algorithm as above. To associate them, we match the scattered peaks with the four original peaks within a matching radius of 3 times the smoothing PSF, $3\times\theta_g$ (see \autoref{subsec:2DK}). For each substructure, we then calculate $90\%$ confidence bounds, which we overlay in \autoref{fig:BS} (green contours).
The Core, our most significant peak,  is detected in 98\% of the bootstrap realizations, and out of those it is the most significant in 91\%, making it very robust. \textcolor{red}{The W substructure is ranked the second-most massive or most massive in 53\% of the realizations in which it is detected, while the NW and NE are ranked as 2nd or 1st in  48\% and 17\%, respectively. 
This demonstrates the ranks of the substructures significance are quite robust. }

We estimate the redshift of each component
as the median photometric redshift of ``green'' galaxies  (CC-selected to represent the cluster population; see \autoref{subsec:clsample}) that lie within $1\arcmin$ from the respective subhalo peak. This substructure lens redshift is used to estimate the respective effective source redshift, $z_\mathrm{s,eff}$. These quantities are listed in \autoref{tab:subhalos} (columns 4--5). According to the peak redshifts, all the substructures are part of the cluster, at $z_\mathrm{l}\approx0.31\pm0.1$.

\subsection{Multi-Halo Mass Modeling}
\label{subsec:2Dhalo}


\begin{deluxetable*}{lrrcccc} 
\tabletypesize{\footnotesize}
\tablecolumns{7}
\tablecaption{\label{tab:nfwMH} Best-Fit NFW Multi-Halo Parameters of the 2D Shear Analysis} 
\tablewidth{0pt} 
\tablehead{ 
\multicolumn{1}{l}{Halo} &
\multicolumn{1}{c}{$X_\mathrm{c}$} &
\multicolumn{1}{c}{$Y_\mathrm{c}$} &
\multicolumn{1}{c}{$R_\mathrm{200c}$} &
\multicolumn{1}{c}{$M_\mathrm{200c}$} &
\multicolumn{1}{c}{$\Delta \mathrm{BIC}$} &
\multicolumn{1}{c}{$\ln{E}$} 
\\ 
\multicolumn{1}{l}{} &
\multicolumn{1}{c}{(arcmin)} &
\multicolumn{1}{c}{(arcmin)} &
\multicolumn{1}{c}{(Mpc)} &
\multicolumn{1}{c}{($10^{15}M_\odot$)} &
\multicolumn{1}{c}{}& 
\multicolumn{1}{c}{} 
}
\startdata 
\sidehead{1-halo model (NFW)}
(1) Core & $0.04\pm0.18$ & $-0.41\pm0.16$  & $2.35\pm0.16$  & $2.06\pm0.42$ & 14 & -6044 \\ 
\\[1mm]
\hline
\sidehead{4-halo model (NFW+3tNFW, $\tau_{{200c}}=1$)}
(1) Core& $-0.32\pm0.18$ & $-0.71\pm0.16$  & $1.69\pm0.26$  & $0.77\pm0.34$ \\ 
(2) W& $1.37\pm0.42$ & $-0.58\pm0.43$  & $1.42\pm0.22$  & $0.45\pm0.20$ \\ 
(3) NE& $-1.62\pm0.33$ & $0.85\pm0.53$  & $1.21\pm0.24$  & $0.28\pm0.16$ \\ 
(4) NW& $0.98\pm0.37$ & $1.89\pm0.48$  & $1.06\pm0.23$  & $0.19\pm0.12$ \\ 
Total &         &           &                   & $1.76\pm0.23$& 0 & $-6034$\\

\enddata
\tablecomments{
An NFW model was fitted to the main Core, and  three tNFW to the others. The concentration parameter was fitted only for the case of 1-halo model, whereas in the 4-halo case the concentrations were set using the mass-concentration relation given by \cite{Bhattacharya2013}. Where values are derived from the posterior distribution of the MCMC sampling,  bi-weight center and scale estimators are reported.
{Columns~(2--3): The centroid positions of each halo, given relative to the cluster center in units of arcminutes.
The centroids are derived from peak detections in the convergence map, and the errors are derived from the scatter of the peaks detected and matched in 500 bootstrapped maps.
}
{Column~(4): Halo size.}
{Column~(5): Halo mass.}
{Column~(6): Relative Bayesian information criteria (BIC) reported for each model. The ratio is given relative to our best model, 4-halo with $\tau_{200c}=1$ which has the lowest BIC.}
{Column~(7): Bayesian evidence reported for each model. Evidence is calculated as the harmonic mean of the likelihood distribution.}
}
\end{deluxetable*}


\begin{figure*}
 \centering
\includegraphics[width=\textwidth]{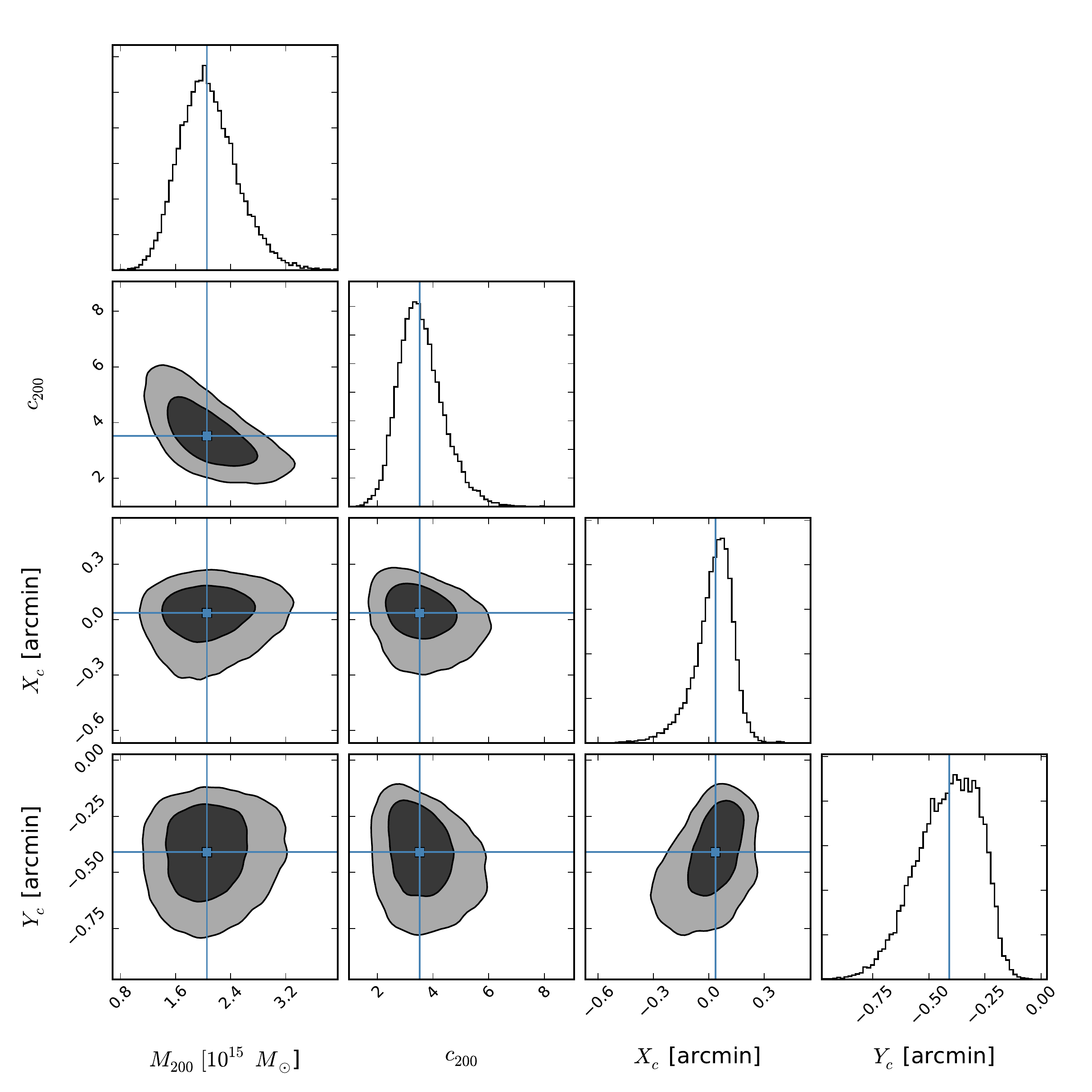}
\caption{The posterior distribution of the 1-halo model parameters, $M_\mathrm{200c}, c_\mathrm{200c}, X_\mathrm{c}, Y_\mathrm{c}$. The contours show the 68\% and 95\% confidence intervals. The solid lines show the marginalized bi-weight mean. }\label{fig:1h_mcmcplots}
\end{figure*} 

\begin{figure*}
 \centering
\includegraphics[width=\textwidth]{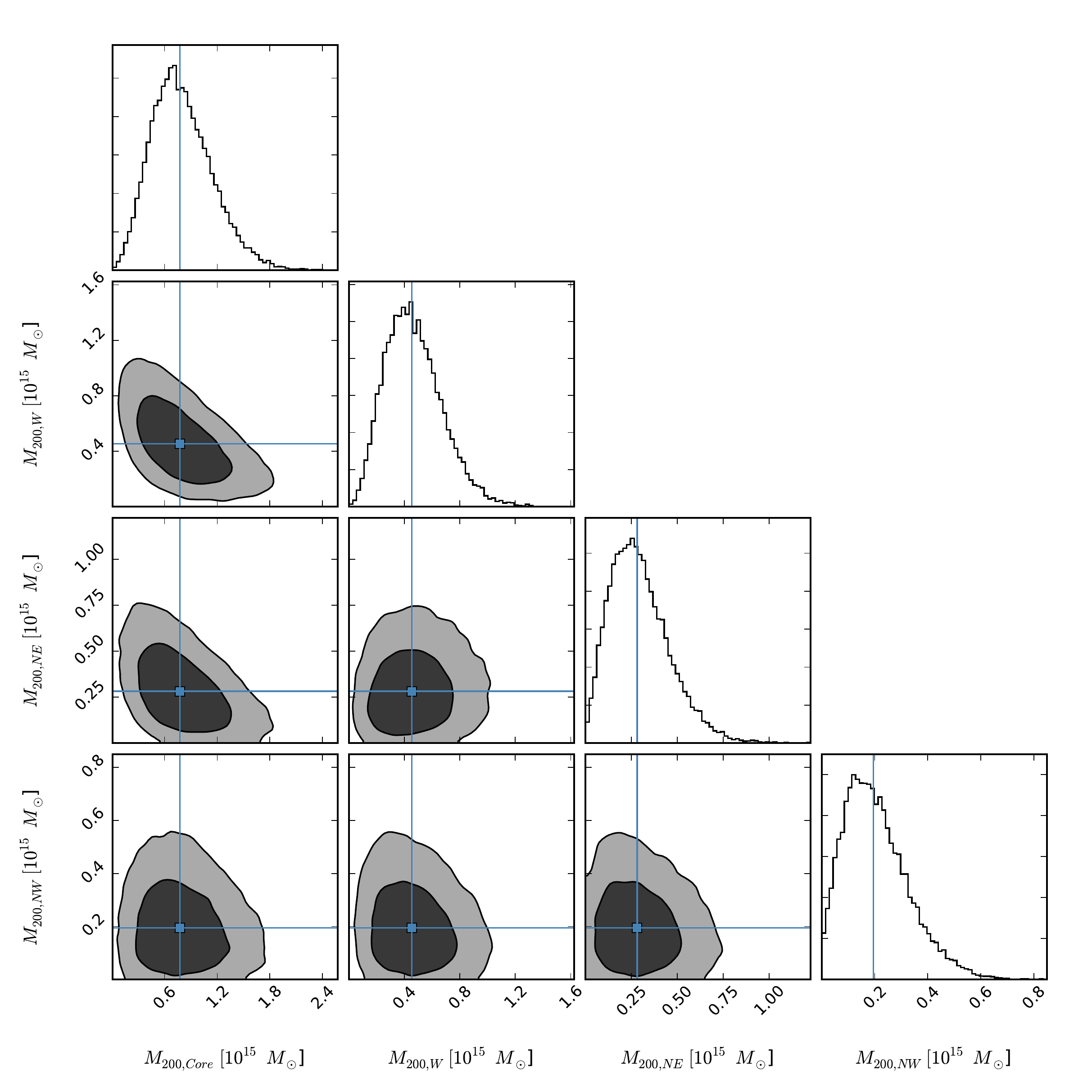}
\caption{The posterior distribution of the 4-halo model parameters, in the case of $\tau_{200c}=1$. The contours show the 68\% and 95\%  confidence intervals. The solid lines show the marginalized bi-weight mean. }\label{fig:4h_mcmcplots}
\end{figure*} 

To estimate individual masses of the structures comprising the cluster, we conduct a two dimensional (2D) shear fitting \citep{Okabe2011,Watanabe2011}, simultaneously modeling the multiple halos detected in the previous section. 
The pixelized distortion map $\langle g(\btheta_m)\rangle$ (\autoref{eq:ggrid}) and its variance map  $\sigma_{g,m}^2$ (\autoref{eq:siggrid}) are constructed by bin averaging ($S=1$, whereas in the map-making we used Gaussian smoothing kernel) in each cell using the statistical weight $w_i$ (\autoref{eq:w}) onto a $208\times208$ grid of independent cells. 

The log-likelihood function of the shear fitting is calculated as
\citep{Umetsu2015}:
\begin{equation}
 \ln \mathcal{L}_g 
= -\frac{1}{2}
 \sum_{m}^{N_{\mathrm pix}}
 \sum_{\alpha=1}^{2}
[g_{\alpha,m}-\hat{g}_{\alpha,m}]
\left({\cal C}^{-1}_g\right)_{m}%
 [g_{\alpha,m}-\hat{g}_{\alpha,m}]
\end{equation}
where  $\hat{g}_{\alpha,m}$
is the theoretical expectation for $g_{\alpha,m}=g_\alpha(\btheta_m)$,
and ${\cal C}_{g,m}$ is the shear covariance matrix, calculated from the
2D pixelized shear (see \autoref{eq:siggrid}) as ${\cal C}_{g,m} = \frac{1}{2} \delta^K_{\alpha\beta}\sigma_{g,\alpha}(\btheta_m)\sigma_{g,\beta}(\btheta_m)$.

To describe the mass distribution, we employ the following models of halo mass density:
\begin{enumerate}
 \item
The universal  \citet[][NFW]{NFW96}  mass density profile, given by the form 
\begin{equation}\label{eq:nfw}
\rho_{\mathrm NFW}(r)=\frac{\rho_{s}}{(r/r_s)(1+r/r_s)^{2}},
\end{equation} 
where $\rho_{s}$ is the characteristic density, and
$r_s$ is the characteristic scale radius at which the logarithmic density slope is isothermal. 
The halo mass $M_{\Delta}$ is  given by integrating the NFW profile (\autoref{eq:nfw})
out to the  radius $r_{\Delta}$, at which the mean density is $\Delta\times\rho_\mathrm{crit}(z_\mathrm{l})$, the critical mass density of the universe at the cluster redshift, expressed as
$ M_{\Delta} \equiv  M(<r_{\Delta})= (4\pi/3)\rho_\mathrm{crit}(z_\mathrm{l})
      \Delta r_{\Delta}^3 $. We use $\Delta=200$ to define the halo
      mass, $ M_\mathrm{200c}$. The degree of concentration is defined as,
      $c_\mathrm{200c}\equiv r_\mathrm{200c}/r_s$, and the characteristic density is
      then given by $\rho_s ={\Delta\times \rho_{\mathrm crit}}/{3}\times{c_\Delta^3}/[{\ln(1+c_\Delta)-c_\Delta/(1+c_\Delta)}]$.

\item
The \cite{Baltz2009} model (aka truncated NFW, or tNFW) is a modification of the NFW model, which suppresses the NFW mass profile beyond a finite radius, $r_t$. This model should better describe merging subhalos, which are not determined by the virial theorem but rather by the strong tidal force of the main cluster. It is expressed as (considering the $n=2$ case)
\begin{equation}
 \rho_{\mathrm tNFW}(r)=\frac{\rho_{s}}{(r/r_s)(1+r/r_s)^{2}}\bigg[\frac{1}{(1+r/r_t)^{2}}\bigg]^2,
\end{equation}
with $r_t=\tau_\mathrm{200c}\times r_\mathrm{200c}$, where $\tau_\mathrm{200c}$ is the ratio of the truncation to the $r_\mathrm{200c}$ radius. The characteristic density, $\rho_s$, is defined as in the NFW profile. Note, the virial mass we derive from this fit, $M_{200c}$, is not strictly the 3D halo mass, as in the case of an NFW profile. We therefore infer the 3D mass inside $r200c$, $M_{\mathrm BMO}(<r200c)$, from the fitted parameter, using Equation~(10) in \cite{Oguri2011}.

\end{enumerate}

We consider two scenarios when modeling this cluster and its substructure: (a) a single mass halo given by the NFW model, not accounting for substructure in this case but modeling the cluster as a whole; 
and (b) four mass halos, with an NFW model for the Core (the most significant peak) and three tNFW models for the other substructures, W, NE and NW. 

For each case, we let different sets of parameters vary: for the 1-halo
solution, (a), we let $M_\mathrm{200c}$ and $c_\mathrm{200c}$ of the mass
profile vary, and also let the center of mass vary ($X_\mathrm{c}, \,Y_\mathrm{c}$) vary, thus having four free parameters to fit for. For the sake of computational efficiency, we set a flat prior on the mass and concentration in the range $0\le M_\mathrm{200c}/\munit\le4$, $0\le c_\mathrm{200c}\le10$.
We test several upper limits to confirm that the choice of priors on the mass and concentration is uninformative. 
In order to be inclusive, yet not to confuse with nearby substructures, we impose a Gaussian prior on the peak centroid. We conservatively set the prior width to three times the scatter on the peak location estimated from the bootstrapped convergence maps (see \autoref{subsec:substr}), $\sigma_{\pi(X_\mathrm{C})}= 0.54\arcmin,\sigma_{\pi(Y_\mathrm{C})}= 0.48\arcmin$.

For the 4-halo solution, (b), we  set the concentration parameter for the main NFW halo and the  three tNFW subhalos based on the known mass-concentration relation taken from \cite{Bhattacharya2013}. \textcolor{red}{We have run several test to validate this assumption (see discussion below). }
Here, since our data are not sensitive enough to fit for the centroids simultaneously, we fix the centroids of the halos to the values derived from the convergence map, and set the errors on their location according to the scatters of the bootstrapped maps (see \autoref{tab:subhalos}). We thus only fit for the halos masses, having four free parameters in total. We set the same flat priors on the masses as above, $0\leqslant M_\mathrm{200c}\leqslant4\times\munit$.
The truncation parameter, $\tau_\mathrm{200c}$, of the three tNFW subhalos depends on the dynamical timescale, the relative strength of the parent to subhalo masses, the distance between the parent and subhalo, etc. It should, in principle, be left as a free parameter. However, given the complexity of this system and the limitations set by our data, we only explore two values, $\tau_\mathrm{200c}=1,3$. The case of $\tau_\mathrm{200c}=3$ is typical for isolated cluster-sized halos, whereas $\tau_\mathrm{200c}=1$ is typical for mildly truncated interacting halos \citep{Takada2003}.  The latter provides a more realistic value in describing the merging subhalos detected in our work (see \autoref{sec:discussion}).
The marginalized parameters and confidence bounds are determined using a Markov chain Monte Carlo (MCMC) method with standard Metropolis-Hastings sampling. We use the bi-weight mean and scale estimators of \cite{Beers1990} to characterize the marginalized one-dimensional posterior distributions.

For each model we calculate the  Bayesian information criteria (BIC) \citep{Schwarz1978}, according to $\mathrm{BIC}=-2\ln\hat{L}+k\ln N$, where $\hat{L}$ is the maximum likelihood estimator, $k$ is the number of free parameters, and $N$ is the number of constraints. We also quote the Bayesian evidence, calculated as the harmonic mean of the likelihood distribution. Both these quantifiers consistently show strong indication that the 4-halo models are preferred over the 1-halo model. 
Both 4-halo models of the two truncation cases, $\tau_\mathrm{200c}=1,3$, gave virtually consistent masses. However, the $\tau_\mathrm{200c}=1$ is marginally preferred over the $\tau_\mathrm{200c}=3$ case, according to their relative BIC, $\mathrm{BIC}_{\tau_\mathrm{200c}=3} - \mathrm{BIC}_{\tau_\mathrm{200c}=1} = 2$.
Thus, we conclude that we are not sensitive enough to fully constrain the tidal truncation radius of the individual merging subhalos but there is weak preference for the $\tau_\mathrm{200c}=1$ case. In what follows, we therefore focus on the $\tau_\mathrm{200c}=1$ results. The marginalized parameter constraints of the 1-halo model and 4-halo ($\tau_\mathrm{200c}=1$) model are summarized in \autoref{tab:nfwMH}. 
The two-dimensional marginalized posterior distributions for all pairs of  parameters of the 1-halo and 4-halo cases are shown in \autoref{fig:1h_mcmcplots} and \autoref{fig:4h_mcmcplots}, respectively.

Not shown in the table, for the 1-halo model we find the concentration parameter to be $c_\mathrm{200c} = 3.5\pm0.8$. This value is consistent with cluster-type halos of this size \citep{Bhattacharya2013}. \textcolor{red}{For the 4-halo solution, we have run several tests letting the concentration parameter vary as well, in order to test the effect of fixing the mass-concentration ratio in our modeling. The masses remained nearly the same, fully consistent within the errors, whereas the concentration was not constrained very well (giving errors of $\Delta c_\mathrm{200c}\lesssim3$, well beyond the range of values between models), and the overall confidence of the model was lower according to $\Delta \mathrm{BIC} = 34$ (w.r.t. fixed concentrations). These low-mass subhalos have scale radii smaller than our resolution scale, so that one cannot constrain the concentrations well.}

\begin{figure}
 \centering
\includegraphics[width=0.48\textwidth]{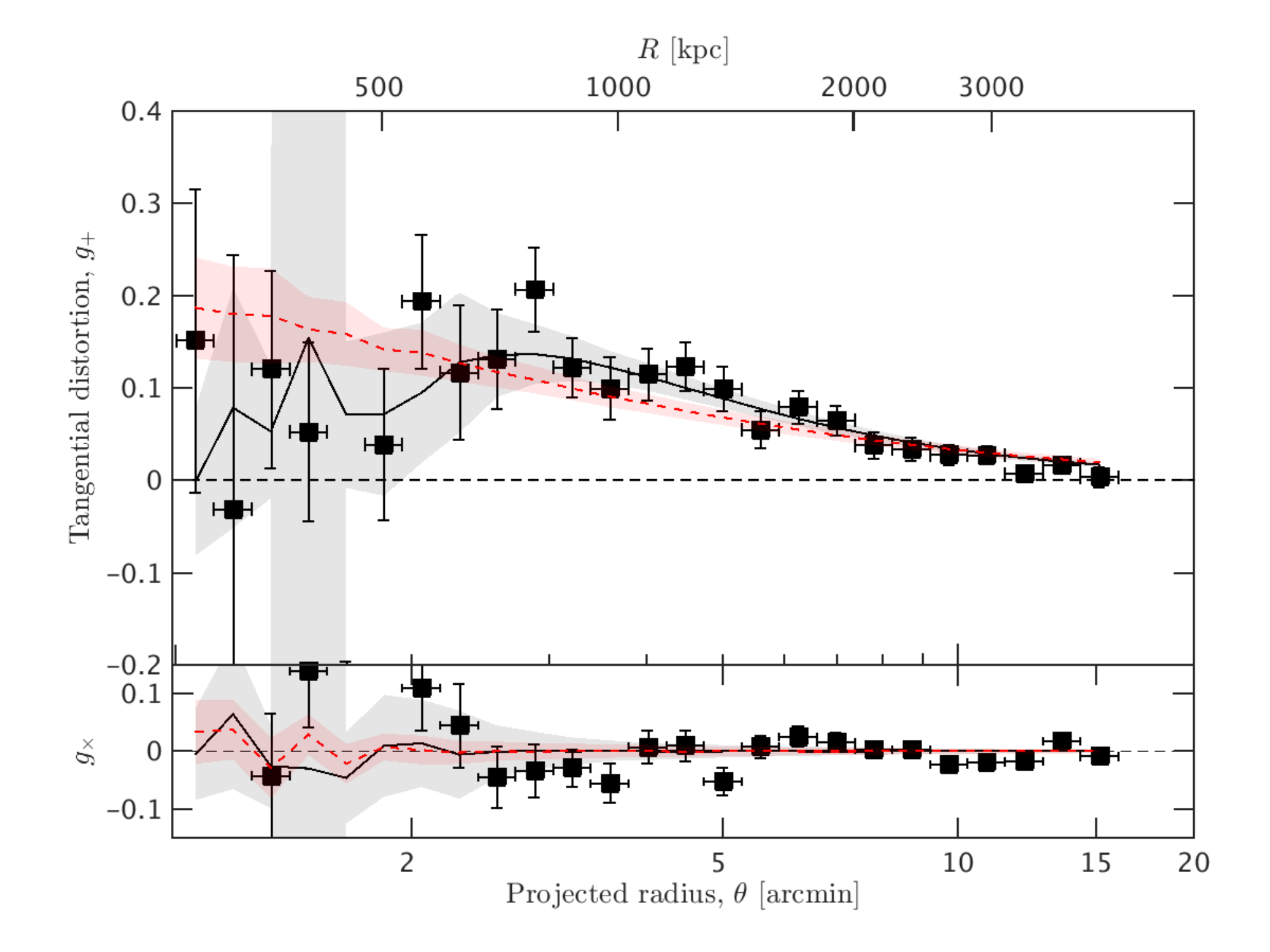}
\caption{Azimuthally-averaged radial profiles of the tangential reduced-shear $g_+$ of the background galaxy sample (black squares), as in \autoref{fig:gt1D}, with comparison to the 1-halo model posterior shear (dashed red line) and its $1\sigma$ scatter (shaded red region) and the 4-halo model (in the case of $\tau_\mathrm{200c}=1$; solid black line and shaded gray region).}
\label{fig:gtcomp}
\end{figure}

We compare the results of the 1-halo and 4-halo model posterior shear profiles with the observed shear profile from \autoref{subsec:gt} in \autoref{fig:gtcomp}. Overall, the 4-halo model indeed describes the observed shear profile better, though it has a broad confidence bounds inside $\theta<2\arcmin$, likely due to the contribution of the substructures near the cluster core.
We further discuss our results below in \autoref{sec:discussion}.

\section{Discussion}
\label{sec:discussion}

In the previous section, we reconstructed the mass distribution of the cluster, detected four distinct substructures near its center, and modeled their masses and locations directly from the shear data. In the following section we discuss our results and their implication on the interpretation of the merger.

\subsection{Total Cluster Mass from Different Methods}
\label{subsec:Mtot}
A2744 is a very massive cluster, as indicated by all the different
methods,  consistently yielding $\sim2\times \munit$. In the single-halo
case, we found a total mass of $M_\mathrm{total}=(2.06\pm0.42)\times\munit$, whereas in
the 4-halo case, the total mass is
$M_\mathrm{total}=(1.76\pm0.23)\times\munit$, calculated from the bi-weight mean posterior of the sum of the four halos. For comparison with M11's projected total mass inside 1.3\,Mpc, from the 1-halo model we find $M_\mathrm{2D}(r<1.3\,\mathrm{Mpc})=(1.65\pm0.23)\times\munit$, in very good agreement with M11 value, $M(r<1.3\,\mathrm{Mpc})=(1.8\pm0.4)\times\munit$.

\subsection{Masses of the Different Cluster Components}
\label{subsec:Msub}
By utilizing our multi-halo modeling technique we were able to estimate the masses of the different components of the cluster. Overall, we find that the most massive halo is the main Core, with a mass of 
$M_\mathrm{200c}=(0.77\pm0.34)\times \munit$ when fitting 4-halos to the reduced shear field. This mass is lower than when fitting just one halo, $M_\mathrm{200c}=(2.06\pm0.42)\times \munit$, demonstrating that in order to best derive the mass of just the Core the different substructures need to be taken into account. We find the second most massive substructure to be the W clump, with a mass of
$M_\mathrm{200c} = (0.45\pm0.20)\times\munit$. Our modeling yields a close mass ratio between these two components,  1.7:1 (but the uncertainty on this value is large: $\pm1.1$).
The other two halos have lower masses, with $M_\mathrm{200c} = (0.28\pm0.16)\times\munit$ for the NE clump and $M_\mathrm{200c} = (0.19\pm0.12)\times\munit$ for the NW clump.

M11 had detected similarly four substructures. However, only three -  {\em their} Core, W and NW peaks - are more closely related to our peaks, albeit at slightly offset locations (which we further discuss below). 
M11 found projected masses of, $M(<250$\,kpc$)=(2.24\pm 0.55),(1.11\pm0.28),(1.15\pm0.23)\times10^{14}\,M_\odot$, for the Core, W and NW peaks, respectively. To compare with M11 values, we compute the total projected mass within 250\,kpc using the 4-halo modeling results. We obtain $M(<250$\,kpc$)=(1.49,1.25,0.76)\times10^{14}\,M_\odot$, with typical errors of $\pm0.35\times10^{14}\,M_\odot$, for the Core, W and NW clumps.  
Overall, we obtain mostly lower but consistent masses -- our Core mass is $(67\pm22)\%$ lower,   our W is $(113\pm42)\%$ higher, and our NW mass is $(66\pm30)\%$ lower.

Most notably, M11 showed the NW to be the second most massive clump, with  a $\sim$2:1 mass ratio to the main Core, whereas in our analysis the W clump is the second most massive instead, and our NW is the least massive clump. However, the determined mass ratios have large uncertainties, given our large errors on the masses for the satellite substructures.

\subsection{Separations between the Gas, Stars and DM in the Substructures}
\label{subsec:Msep}


\begin{deluxetable}{lccc}
\tabletypesize{\footnotesize}
\tablecolumns{4} 
\tablecaption{ \label{tab:offsets} Measured Offsets of DM halos}  
\tablehead{ 
 \multicolumn{1}{c}{Halo} &
 \multicolumn{1}{c}{nearest BCG} &
 \multicolumn{1}{c}{M11's DM halos} &
 \multicolumn{1}{c}{X-ray feature} 
\\
 \colhead{} & 
 \colhead{($\arcsec$)} &
 \colhead{($\arcsec$)} &
 \colhead{($\arcsec$)} 
} 
\startdata  

Core & $6^{+24}_{-1}$ & $16^{+19}_{-8}$ & $13^{+24}_{-6}$ \\ 
W & $72^{+34}_{-53}$ & $80^{+34}_{-54}$ & $91^{+66}_{-27}$ \\ 
NE & $-$ & $24^{+64}_{-6}$ & $-$ \\ 
NW & $68^{+25}_{-42}$ & $69^{+32}_{-63}$ & $87^{+34}_{-28}$ \\ 
\enddata 
\tablecomments{
Errors are estimated as 90\% confidence bounds from the bootstrapped map offsets in each case, as in \autoref{fig:BS}.
Column~(2): Offset is given relative to BCG in the nearest clump of galaxies, used as a tracer of light.
Column~(3): Offset is given relative to published peak locations in M11.
Column~(4): Offset is given relative to nearest X-ray feature - southern cool remnant core for Core, main X-ray emission for the W, and NW interloper for NW subhalo.
}
\end{deluxetable}

The different substructures identified in \autoref{subsec:substr} show some remarkable offsets from both the gas and galaxies in this complex merger. We summarize these offsets in \autoref{tab:offsets} and discuss them below.
We confirm the most massive structure (Core) to be located near ($16^{+19}_{-8}\arcsec$, quoting 90\% CL) the designated Core in M11, and north of where \cite{Owers2011} noted an X-ray cool southern remnant core ($13^{+24}_{-6}\arcsec$).  It is also associated with the BCGs in this location, at about $6^{+24}_{-1}\arcsec$ away, within the confidence bound (green contour in \autoref{fig:BS}). 

As noted above, we find in our analysis the second most massive peak to be the W clump, but surprisingly, to be separated from any BCG. The nearest BCG is  $72^{+34}_{-53}\arcsec$ further west. We note that the W substructure is highly extended southwest, indicating a possible secondary peak closer to the BCG. M11's W clump is much closer to the BCG in that location, and so is $80^{+34}_{-54}\arcsec$ away from our W peak. Although the nearest X-ray emission peak is $91^{+66}_{-27}\arcsec$ further east, there is a clear extended X-ray emission pointing in its direction (\autoref{fig:BS}). 

M11's second-most significant peak is located NW of the core, associated with the second-most luminous clump of galaxies. We, on the other hand, find an extended mass distribution (clump NW) with its peak  $69^{+32}_{-63}\arcsec$ north of M11's NW location, and about $87^{+34}_{-28}\arcsec$ east of the X-ray interloper peak. 
It is also  $68^{+25}_{-42}\arcsec$ north of the closest BCG in the northwest clump of galaxies, and $58^{+45}_{-14}\arcsec$ west of the closest BCG in the north clump of galaxies, so it seems to lie just between these two BCGs.

Finally, the location of the NE clump is detected close ($24^{+64}_{-6}\arcsec$ south) to where M11 show a low-significance clump (not listed in their paper). The X-ray emission is extended toward the NE, though no X-ray peak is detected next to the NE substructure. The NE clump has the lowest significance ($\sigma_\kappa=4.7$) of all four peaks, and it appears to be extended and bimodal. In fact, the convergence map shows another peak which is part of the same extended NE distribution, but at a lower significance ($\sigma_\kappa=4.48$), and was therefore not included in our modeling. Although there are no clear BCGs in this location, there are several faint cluster members ($\RC\approx23$ mag).

To demonstrate the different levels of mass-galaxy partition, we investigate the mass-to-light ratio inside a small aperture around each mass peak, $\sim250\,$kpc. We use the masses $M_\mathrm{2D}(<250\mathrm{kpc})$ derived from the 4-halo model (see \autoref{subsec:Msub}). We estimate the total luminosity inside the same aperture using K-corrected $\RC$-band luminosities of the cluster member galaxies (the ``green'' sample, see \autoref{subsec:clsample}). We find the projected mass-to-light ratios to be:
$M_\mathrm{2D}(<250\,\mathrm{kpc})/L_{R_{\mathrm C}}(<250\,\mathrm{kpc}) = (103\pm25, 584\pm162, 366\pm134, 153\pm 63)\times\,M_\odot/L_{\odot}$ for the Core, W, NE and NW clumps, respectively. In comparison, the projected mass-to-light ratio of the entire cluster, calculated using the 1-halo mass model, is $M_\mathrm{2D}(<r_\mathrm{200c})/L_{R_{\mathrm C}}(<r_\mathrm{200c}) = (241\pm38)\times\,M_\odot/L_{\odot}$. This value is typical for cluster-sized halos. The value we measure for the W peak is about $2.4\pm0.8$ times higher than that typical value, and hence reveals that this clump is deficient of galaxies. 
Finally, the Core shows a rather low mass-to-light ratio, $0.4\pm0.1$ that of the entire cluster. This stems from the contribution of the two BCGs which are inside the Core inner $250$~kpc. Lower values near cluster centers have also been indicated previously \citep{Medezinski2010}.

\subsection{Understanding the Merger Scenario}
\label{subsec:Merger}
\begin{figure}
 \centering
\includegraphics[width=0.48\textwidth]{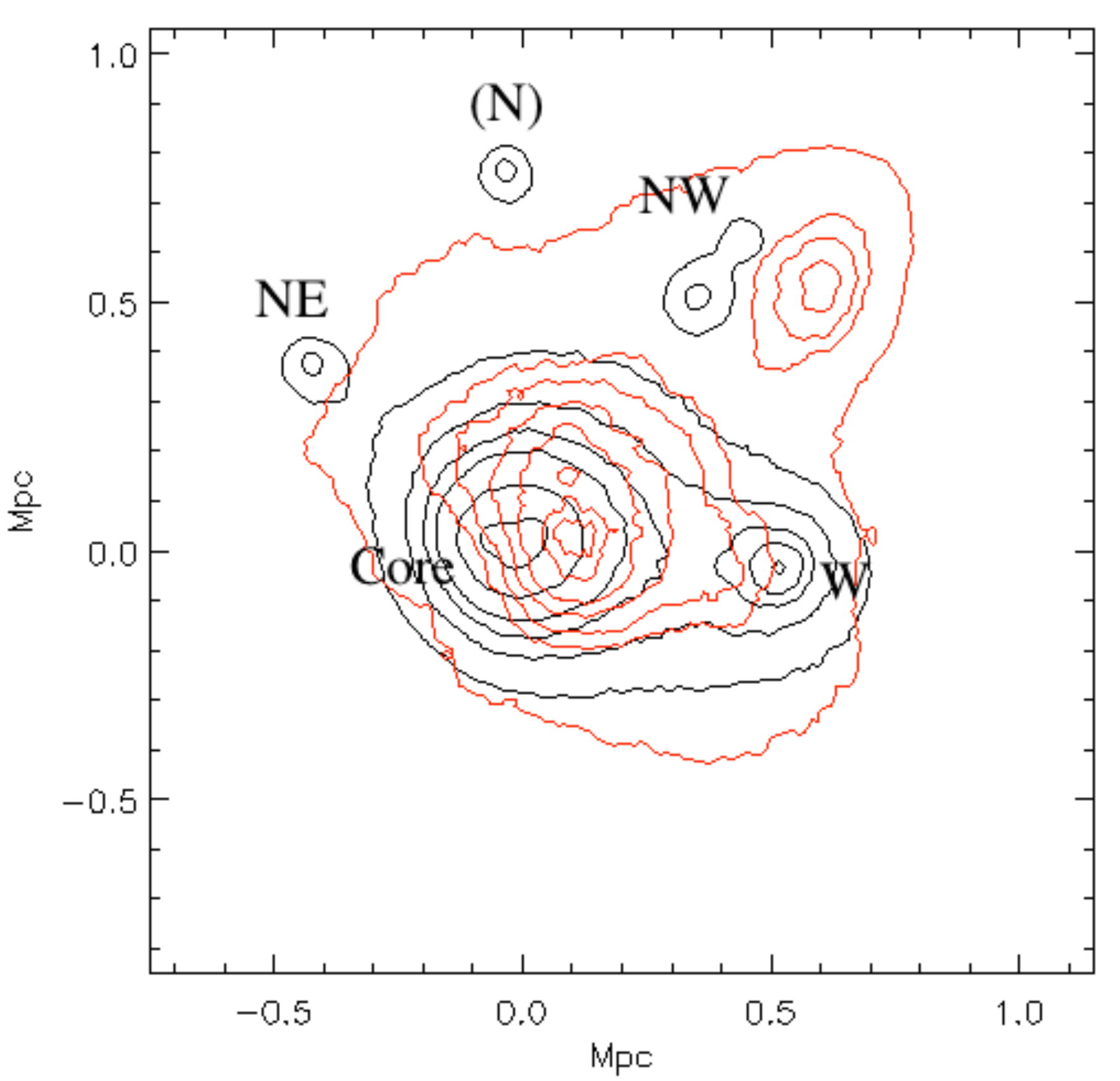}
\caption{Contours of dark matter surface density (black) and X-ray emission (red)
based on our $N$-body/hydrodynamical (FLASH) simulations of merging clusters.
The projection angles were chosen to produce an image which 
resembles the observed morphology of A2744.
See text for details.
}
\label{fig:HydroSimContour}
\end{figure}

Multifrequency observations show that A2744 is a very complex merging
system (see our discussion above).  Our WL reconstruction suggests
that this complicated morphology is a result of multiple mergers due
to infalling subclusters. 
The details of the merging event in A2744, however, are not clear.
The complex morphology of A2744 can only be explained 
with dedicated $N$-body/hydrodynamical simulations, 
which is out of the scope of this paper. 
Here we aim to provide a merely qualitative analysis based 
on a superposition of several binary mergers from our existing archive of well established, self-consistent $N$-body/hydrodynamical 
simulations, which rely on the publicly available code FLASH \citep{Fryxell2000}.  
Our simulation assumes initially spherical clusters with gas following 
a non-isothermal $\beta$ model in hydrostatic equilibrium with a truncated NFW 
model for the dark matter. We used a box size of 13.3 Mpc on a side, and 
the highest resolution (cell size) was of 12.7 kpc.
Our simulations were semi-adiabatic in the sense that only adiabatic processes and 
shock heating were included. A detailed description of the setup for our simulations 
and a discussion of the input parameters can be found in \cite{Molnar2012}.  

Different scenarios have been suggested recently for this system,
although none of them can explain all the observed features
(M11; \citet{Owers2011}).  These interpretations agree that the
dominant X-ray emission
is a result of a major merger along the north-south axis, and that its
offset from the DM mass peaks is due to merging, but they differ on
which component is the massive cluster core.  \cite{Owers2011},
based on X-ray and galaxy velocity distributions, suggest
that the main cluster is the northern component (identified as a peak
in the galaxy surface density distribution to the north of the main X-ray
peak; see left panel in Figure~13 of \citet{Owers2011}), and that the
southern component (close to the southern X-ray compact cool core) is
the infalling, less massive bullet subcluster. Based on a combined SL+WL analysis, M11 proposed that the southern component, which they find as the more massive one, is in fact the main core. 
Our WL analysis confirms M11's suggestion that the most massive 
component of A2744 is the southern core (see \autoref{fig:color} and \autoref{fig:BS}  and \autoref{tab:nfwMH}). 
However, our WL analysis suggests that the direction of the major 
merger (involving the two most massive components -- the Core and W) 
is along the east--west direction, as opposed to M11's results which 
suggest a major merging event in the southeast--northwest 
direction (their Core and NW1 and NW2 components).

A likely explanation for our detected NE and W mass components is that they have 
passed through 
the core moving towards east and west, respectively, and lost most of their gas 
due to ram pressure from the more massive main halo.
This picture explains the X-ray tails toward the northeast and southwest 
seen in the X-ray map (\autoref{fig:BS}). 
In order to explain the X-ray morphology extended to the north,
we need to assume a south--north merger as well. 
Although the northern mass peak we found (see \autoref{fig:BS}) is not very significant ($\sigma_\kappa=4.0$) and was therefore not included in our modeling, we include it here in our scenario, and assume it is much less massive than the Core. We label it accordingly as (N), to differ it from the other significant halos that were included in the mass modeling.
If such a subhalo passed the Core from the east, moving from the south to the north, it can push the gas of the main cluster to the northwest.

It is much more difficult to explain the X-ray interloper located 
to the northwest.  
M11 suggest that the interloper is a result of a slingshot 
due to a multiple merger event. According to M11's scenario, 
the mass peaks closest to the X-ray interloper (their NW1 and NW2), 
passed through the main cluster from 
the southeast to northwest (see Figure 9 of M11).  
As the subclusters climb out of the potential well of the main cluster 
moving to the northwest, both the ram pressure and the gravitational 
potential (which was temporarily enhanced by the NE and W subhalos 
passing through the core) drop, and the remaining gas components 
of NW1 and NW2 suffer a slingshot effect and get ahead of their DM centers.  
However, hydrodynamical simulations \citep{Molnar2012} suggest that 
at this large distance from the Core of the main cluster ($\sim 600$\,kpc) 
the slingshot gas should have already fallen back to its DM halo (NW1+NW2), 
making the slingshot explanation of M11 unlikely.

Similar to M11, we find a significant mass concentration to the
northwest, clump NW, closest to the interloper ($\sim\,$393\,kpc).  
Our NW component is offset north of M11's NW component, but at about the same distance from the interloper.  
Even in this setup, the slingshot effect is not probable, as the offset between our NW clump and the interloper is opposite to the NW-Core axis. 
Our NW component is elongated in the southeast--northwest
direction, which suggests that it may comprise of two unresolved, 
less massive halos. Under this assumption, we suggest an 
alternative scenario, where the two subclusters NW1 and NW2 collide before they reach the main cluster. The less massive of the two subclusters is infalling from the northwest close to the LOS through the larger one, displacing the gas from the potential well of the subclusters. As a result of the LOS projection, the X-ray peak appears to be located  west of the two DM peaks.
A different scenario to be considered is where the interloper gas may be tidally stripped from the W substructure by the northern mass peak (N). However, the large
physical distance of the X-ray peak from both the W and the northern mass components makes this scenario highly unlikely, and in the following we pursue the infalling NW scenario to explain the interloper.

\begin{deluxetable}{lcccc}
\tabletypesize{\footnotesize}
\tablecolumns{5} 
\tablecaption{ \label{tab:sims} Simulations of the merging substructures}  
\tablehead{ 
 \multicolumn{1}{l}{Merger} &
 \multicolumn{1}{c}{$M_\mathrm{tot}$} &
 \multicolumn{1}{c}{Mass Ratio} &
 \multicolumn{1}{c}{$P$}  &
 \multicolumn{1}{c}{$V_0$} 
\\
 \colhead{} & 
 \colhead{($\munit$)} &
 \colhead{} &
 \colhead{(Mpc)} &
 \colhead{(km s$^{-1}$)} 
} 
\startdata  
W-Core  & 1.4 & 1.8 & 0.2  & 4500\\
NE-Core & 0.8 & 3   & 0.15 & 4500\\
(N)-Core & 1.2 & 5.2 & 0.15 & 4000\\
NW1-NW2 & 0.23& 6.67& 0.15 & 4500\\
\enddata 
\tablecomments{
Column~(2): Total virial mass of the Core and satellite halo.
Column~(3): Mass ratio of the Core to satellite.
Column~(4): Impact parameter.
Column~(5): Infall velocity.
}
\end{deluxetable}
In order to demonstrate the validity of our suggested scenario, we superimpose four separate binary simulations from our archive, three to explain the main X-ray emission and one to explain the NW interloper (see \autoref{tab:sims}). 
The first simulation is used to represent the east-to-west merger, associated with component W, with a mass ratio of 1:1.8 (close to the observed value), a total mass of $1.4\times\munit$, 
an impact parameter of 0.2\,Mpc and an infall velocity of 4500\,km\,s$^{-1}$.
The second simulation represents the west-to-east merger, associated with component NE, assuming a mass ratio of 1:3, a total mass of $0.8\times\munit$, an 
impact parameter of 0.15\,Mpc and an infall velocity of 4500\,km\,s$^{-1}$. 
The third simulation is used to represent the south-to-north merger, associated with component (N),
assuming a mass ratio of 1:5.2, a total mass of $1.2\times\munit$, an 
impact parameter of 0.15\,Mpc and an infall velocity of 4000\,km\,s$^{-1}$. 
The fourth merging simulation is used to represent the NW minor merger of two smaller subhalos having a
total mass of $0.23\times\munit$, a mass ratio of 1:6.67, an impact parameter of
0.15\,Mpc and an infall velocity of 4500\,km\,s$^{-1}$. 
The masses and concentration parameters of the Core
in our binary simulations lie in the range $0.6\le M_\mathrm{vir}/\munit\le1.0$ 
and $5\le c_\mathrm{vir}\le8$, as allowed by our mass modeling.
We choose the outputs (epochs) and the viewing angles of each simulation to qualitatively reproduce the observed surface mass density distribution and X-ray morphology. 
We assumed that the final Core component in the different merging simulations is the same, but viewed at different epochs, thus represented by different ``effective masses''.
Therefore, when stacking the binary simulations we rescale the mass of the Cores so that it would be consistent with the observed mass ($\sum_{i=1,3} w_i M_i= 0.8 \times\munit$). We used the same weights ($w_i$) to rescale the associated Core X-ray emission accordingly.
In Figure~\ref{fig:HydroSimContour} we show the contours of the DM surface density (black) and X-ray emission (red) of all four simulations superimposed, while normalizing the mass of the Core.
We associate the mass peaks of our simulated image with those detected in our WL reconstruction by labeling them  Core, NE, W, NW  and (N).

The morphology of the main components resembles the observed features of A2744.
The mass peaks of our simulations coincide with the locations of the 
observed Core, the W component and the non-significant peak (N). 
The extension of the main X-ray emission to the west towards the W component and to the north are clearly visible and resemble the {\it Chandra} observations.
Our simulation also reproduces the mass peak elongated in 
the southeast--northwest direction similar to our NW component. 
The associated X-ray interloper emission 
is located just west of the NW mass peaks, as observed, 
although the offset in the simulated image is somewhat smaller , $\sim200$\,kpc.  
However, neither our binary merger nor M11's slingshot scenario can 
explain the northern edge of the interloper (which is interpreted as a 
cold front by \cite{Owers2011}) and all other observations consistently. 
Moreover, systematics in the surface mass density reconstruction 
(see \autoref{subsec:substr}) leading to large uncertainties in the positions 
of our less massive components make it difficult to accurately constrain 
the level of offset between the NW mass peak and the X-ray interloper, 
and thus to strongly distinguish between suggested merger scenarios of this component.


\section{Summary and Conclusions}
\label{sec:summary}
We have presented an improved WL analysis of the merging cluster A2744, using new deep $BRz'$ imaging from Subaru/Suprime-Cam. We reconstruct the total mass distribution out to about twice the virial radius of the cluster, $\theta=18\arcmin$ (approximately 5\,Mpc at $z=0.308$).
The deep multi-band imaging allows us to isolate background galaxies, and thus perform a robust WL analysis, removing contamination and dilution of the WL signal which is a dominant source of systematics in cluster WL analyses, particularly near cluster centers. Obtaining an undiluted mass map, we detect four distinct substructures in the cluster central region with higher confidence than before, which we label Core, W, NE and NW, with significance of $12\sigma_\kappa$, $7.9\sigma_\kappa$, $4.7\sigma_\kappa$ and  $7\sigma_\kappa$, respectively. 

We constrain the total mass of this massive cluster to be $M_\mathrm{200c} = (2.06\pm0.42)\times 10^{15}\,M_{\odot}$. By simultaneously fitting an NFW halo and three  tNFW halos to the four detected substructures, we constrain the substructures masses to be $M_\mathrm{200c} = (0.77\pm0.34)\times\munit$ for the southern Core, $M_\mathrm{200c} = (0.45\pm0.20)\times\munit$ for the W substructure, $M_\mathrm{200c} = (0.28\pm0.16)\times\munit$ for the NE substructure and $M_\mathrm{200c} = (0.19\pm0.12)\times\munit$ for the NW substructure. 

Although on larger scales ($\gtrsim1$\,Mpc) the gas, mass and galaxies
seem to be distributed similarly, on smaller scales ($\lesssim200$\,kpc)
there are significant inconsistencies between the positions of gas and the total mass, but
also between the total mass and the galaxies. The massive main core does
appear close to the southern X-ray core, and close to the BCGs. Our second most massive peak (W), on the other hand, is offset $72^{+34}_{-53}\arcsec$ east of the western BCG, toward the cluster core, and appears to be stripped of
both gas and BCGs (though some lower brightness cluster members are detected), with $M_\mathrm{2D}(<250\,\mathrm{kpc})/L_{R_{\mathrm C}}(<250\,\mathrm{kpc}) =(584\pm162)\times\,M_\odot/L_{\odot}$. Such  an offset between DM and galaxies could imply that the DM is not as collisionless as the galaxies.
However, given the large uncertainties on the substructure masses and their level of deficiency/offset relative to the galaxies, we cannot put a meaningful lower bound on the DM cross section for collision at this point.

Based on the 
mass ratios we found, we argue that a major merger may have occurred along the 
west--east axis, with another south-north merger perturbing the main X-ray 
emission to the north. The X-ray interloper to the NW is harder to explain 
given our observations. We find it is unlikely to be the remnant of a slingshot 
effect as suggested by \cite{Owers2011} and M11, given the large distances we 
find between the different mass components. We offer an alternative explanation, 
where the NW clump is infalling into the cluster from the northwest, and is 
comprised of two subhalos that went through a separate third merger close to the LOS, causing the gas to be stripped out and to appear as separate to both. To support our hypothesis, we provide actual $N$-body/hydrodynamical simulations of binary mergers with mass ratios close to those derived from our WL reconstruction that nicely reproduce many of the observed features.
We used a superposition of independent binary mergers, yet
it is clear that the multiple mergers have interacted with each other. 
Thus, one simulation of several mergers of more than two components is needed in order to fully explain the observed features of A2744,
for which our analysis here may serve as a starting point
(Molnar et al., in prep.).

Although consistent within $1-2\sigma$ with the subhalos masses and positions of M11, we find some differences, most notably in the location of the W and NW substructures. We conclude based on the careful systematics analysis in \autoref{subsec:substr} and bootstrap tests, that the major differences between our WL analyses is mostly due to biased source selection. Another major difference is not including SL constraints in our analysis as was done in M11 in the center of the field. Adding SL constraints, with their possibly underestimated errors, largely dominates over WL constraints, possibly giving more weight to the central mass core. Our WL analysis is improved compared to M11's wide-field WL analysis in terms of seeing (FWHM$_{R_\mathrm{C}}=1.16\arcsec$ versus M11's FWHM$_{i'}=1.42\arcsec$), depth (${R_\mathrm{C}}\leq26.83$ versus M11's ${R_\mathrm{C}}\leq26.31$) and careful background selection. However, the quality of our Subaru imaging here is still poor compared with the typical level achievable with Subaru, reaching FWHM$_{R_\mathrm{C}}=0.6\arcsec$ as in our previous WL studies \citep{Medezinski2010,Umetsu2014}.

Including {\it HST} WL and SL analysis using the new, deep FF observations will further improve the constraints on the substructures masses and locations, which we plan to pursue in a subsequent paper (Merten et al. in prep). However, the current FOV of the {\it HST} FF imaging covers the inner $\sim2\arcmin\times2\arcmin$, only probing the Core and part of the NW substructure region, and not fully covering the other satellite substructures. Furthermore, combining  observations of different instruments and depths (as Subaru and {\it HST}) may further lead to systematics, necessitating careful weighting of the different regions of interest.  An optimal approach would be to tile over the inner $\lesssim5\arcmin$ with {\it HST}/ACS with several pointings for full coverage of the four substructures identified, which would provide an order of magnitude improvement in the spatial resolution of the WL mass map.


\acknowledgments
We acknowledge useful discussions with Dan Coe, Orly Gnat, Barak Zackay and Assaf Horesh. We acknowledge useful comments on the manuscript from Massimo Meneghetti, Jessica Krick and Tom Broadhurst. We acknowledge the use of the Astronomical Matlab Packages by \cite{Ofek2014}.  We acknowledge the use of python codes by \cite{Foreman-Mackey2014}.
We thank Nick Kaiser for making the IMCAT package publicly available.  
This research project was supported by the Israeli Centers of Excellence (I-CORE) program, (center no. 1829/12).
EM was partly supported by NASA grant {\it HST}-GO-12065.01-A.
KU acknowledges partial support from 
the Ministry of Science and Technology of Taiwan under the grant
MOST 103-2112-M-001-030-MY3.
MN acknowledges support from PRIN-INAF 2014.
This work was supported by ``World Premier International Research Center
Initiative (WPI Initiative)'' and the Funds for the Development of Human
Resources in Science and Technology under MEXT, Japan, and  Core Research for 
Energetic Universe in Hiroshima University (the MEXT program for promoting the 
enhancement of research universities, Japan). NO is supported by a Grant-in-Aid 
from the Ministry of Education, Culture, Sports, Science, and Technology
of Japan  (26800097).
The $N$-body/hydrodynamical code FLASH used in this work was in part 
developed by the DOE-supported ASC/Alliance Center for Astrophysical 
Thermonuclear Flashes at the University of Chicago. We thank the 
Theoretical Institute for Advanced Research in Astrophysics (TIARA), 
Academia Sinica, for allowing us to use their high performance computer
facility for our simulations.
JM is supported by the People Programme (Marie
Curie Actions) of the European Union's Seventh Framework
Programme (FP7/2007-2013) under REA grant agreement
number 627288.

\bibliography{Elinor}

\end{document}